\documentclass[twocolumn,prd,aps,tightenlines,floats,floatfix,preprintnumbers,nofootinbib,eqsecnum]{revtex4-2}

\def\sp{\kern +3pt}
\def\sm{\kern -7pt}
\def\spQ{\kern +6pt}

\def\bea{\begin{eqnarray}}
\def\eea{\end{eqnarray}}

\def\sfrac#1#2{{\textstyle \frac{#1}{#2}}}

\def\rmk{\rm k}

\def\be{\begin{equation}}
\def\ee{\end{equation}}
\def\ba{\begin{eqnarray}}
\def\ea{\end{eqnarray}}

\usepackage{graphics}
\usepackage{graphicx}
\usepackage{epsf}
\usepackage{amsmath}
\usepackage{amssymb}
\usepackage{bbold}

\usepackage{xcolor}

\usepackage{bbold}
\usepackage{color}
\usepackage[normalem]{ulem}
\usepackage{graphicx,color}
\usepackage{amsmath}
\usepackage{xcolor}

\usepackage{fnbreak}

\begin{document}

\phantom{0}
\vspace{-0.2in}
\hspace{5.5in}

\preprint{{\bf LFTC-23-5/78}} 

\vspace{-1in}

\title
{\bf 
Meson cloud contributions to the 
Dalitz decays \\ of decuplet  to octet baryons}
\author{G.~Ramalho$^1$ and K.~Tsushima$^2$}
\vspace{-0.1in}

\affiliation{$^1$Department of Physics and OMEG Institute, Soongsil University, \\
Seoul 06978, Republic of Korea
\vspace{-0.15in}}
\affiliation{$^2$Laborat\'orio de 
F\'{i}sica Te\'orica e Computacional -- LFTC,
Universidade Cidade de  S\~ao Paulo,  
01506-000,  S\~ao Paulo, SP, Brazil}

\vspace{0.2in}
\date{\today}

\phantom{0}

\begin{abstract}
We study the role of the meson cloud on the electromagnetic
transitions from decuplet  ($B'$) to octet ($B$) baryons
in terms of the squared four-momentum transfer $q^2$.
In the quark model framework, the meson cloud dressing of the quark cores
gives important contributions to the $\gamma^\ast N \to \Delta(1232)$
transition form factors.
In the present work, we estimate the meson cloud contributions of 
all decuplet  to octet baryon transitions
($\gamma^\ast B \to B'$ or $B' \to \gamma^\ast B$). 
Models that combine valence quark effects 
with pion and kaon cloud dressing provide 
a fair description of the radiative decays of decuplet  to octet baryons,
namely the $\Sigma^0(1385) \to \gamma \Lambda (1116)$
and $\Sigma^+(1385) \to \gamma \Sigma^+ (1193)$ decays.
Previous studies indicated the relevance
of the pion cloud effects on the  $B^\prime \to \gamma^\ast B$ 
transition but also suggested that
the kaon cloud contributions may  be important in the timelike region.
We combine then the contributions of the bare core, estimated by a covariant quark model,
with $q^2$-dependent contributions of pion and kaon clouds.
We use the framework to calculate the Dalitz decay rates 
and the Dalitz decay widths of decuplet baryons in 
octet baryons with di-electrons ($B' \to e^+ e^- B$) 
or di-muons ($B' \to \mu^+ \mu^- B$).
We conclude, based on the magnitude of our 
results, that most estimates of the $B' \to e^+ e^- B$ Dalitz decay widths 
may be tested at HADES and PANDA (GSI) in a near future.
We discuss also the possibility of measuring 
the $\Delta (1232) \to \mu^+ \mu^- N$ and $\Sigma^0 (1385) \to \mu^+ \mu^- \Lambda (1116)$ 
decay widths in some facilities, based on the estimated branching ratios.
\end{abstract}

\vspace*{0.9in}  
\maketitle

\section{Introduction}
\label{secIntro}

In the last decades there was a significant progress 
in the study of the electromagnetic structure of baryons  
using electron-baryon scattering ($e B \to e B'$) experiments,
where $B$ and $B'$ are baryons.
In these experiments, the baryon target is probed by a spacelike 
photon ($\gamma^\ast B \to B'$), allowing 
the measurement of transition form factors
in terms of $Q^2 \ge 0$, where $Q^2=-q^2$ and $q$,  
the four-momentum transfer~\cite{Aznauryan12,PPNP2023,Burkert04}.
These studies are presently restricted to 
nucleon ($N$) targets and nucleon excitations ($N^\ast$) in the final state
($\gamma^\ast N \to N^\ast$ transition)~\cite{Aznauryan12,PPNP2023,NSTAR,Burkert04,Drechsel07,Mokeev22a}.
Unfortunately, the electron-nucleon scattering method 
cannot be extended from nucleons to other baryons
due to the short lifetime of the baryon 
states~\cite{Aznauryan12,Briscoe15,Afanasev12}.

The study of the $\gamma^\ast B \to B'$ transition
with timelike photons ($Q^2 < 0$)
is more complex experimentally, but can be accessed 
through the decay $B' \to \gamma^\ast B$, where 
the photon is identified by the decay into a dilepton pair ($\ell^+ \ell^-$),
where $\ell = e, \mu$
(electron or muon)~\cite{HADES17,HADES21a,ColeTL,Ramstein18a,Ramstein19,Timelike2,DecupletDalitz}.
The state $B'$ can be generated by scattering of mesons on 
baryons and by proton-proton or proton-nuclei 
scattering~\cite{Briscoe15,HADES17,HADES20a,HADES17b,HADES20d}.
The reaction $B' \to \ell^+ \ell^- B$ defines 
the Dalitz decay of the baryon $B'$ into 
the baryon $B$ and dileptons.
Different from the traditional electron-nucleon scattering, 
at HADES and PANDA, one can probe the electromagnetic structure of the
baryons in the kinematic region $4 m_\ell^2 \le q^2 \le (M_{B'} -M_B)^2$,
where $M_{B'}$ and $M_B$ are respectively the initial and final baryon masses,
and $m_\ell$ is the mass of the electron or the muon~\cite{Lalik19,Ramstein18a,Ramstein19}. 
The baryon dilepton decays provide then information about 
the electromagnetic structure of the baryons in the timelike 
region ($Q^2 = -q^2 < 0$)~\cite{HADES17,HADES21a,Briscoe15,HADES14,ColeTL,Ramstein18a,Ramstein19,Lalik19,Timelike2,DecupletDalitz}.
Experiments in the timelike region
allow also the possibility
of studying the electromagnetic structure of baryons 
with strange quarks (hyperons)~\cite{HADES21a,DecupletDalitz,Lalik19}.

In the present work, we restrict our study to {\it direct}
electromagnetic decays, excluding weak radiative 
decays that involve weak transitions (decay of strange quarks)
before the final electromagnetic decay.
Examples of these processes are the observed decays
$\Lambda \to \gamma n$, $\Sigma^+ \to \gamma p$, 
$\Xi^0 \to \gamma \Lambda$
and $\Xi^- \to \gamma \Sigma^-$, as well as the
Dalitz decays $\Sigma^+ \to \mu^+ \mu^- p$
and  $\Xi^0 \to e^+ e^- \Lambda$~\cite{PDG22}.

The first measured Dalitz decay width of a nucleon resonance,
the $\Delta (1232)$ decay into di-electrons was performed recently at
HADES~\cite{HADES17,Timelike,Timelike2,Dohrmann10}.
Measurements of Dalitz decays of other 
nucleon resonances are in progress at 
HADES~\cite{HADES22a,ColeTL,Ramstein18a,Ramstein19,N1520TL,N1535TL,Zetenyi21a,HADES20a}.
Recent feasibility studies on HADES indicate that 
measurements of hyperon Dalitz decay rates, 
including $\Sigma (1385)$, $\Lambda (1404)$ and $\Lambda (1520)$ 
into electron-positron pairs are expected at GSI in a near 
future~\cite{HADES21a,HADES20b,HADES20c,HADES20d,Lalik19}. 
The production of hyperons are possible 
at HADES and PANDA due to the large acceptance 
and excellent particle identification, including dileptons in 
the final state~\cite{Salabura13,Lalik19,Rathod20a,Singh16}.
There is also the expectation that measurements 
of baryon Dalitz decays into di-electrons and di-muons
may be observed in PANDA, BESIII and LHCb experiments 
in the following years~\cite{BESIIIreview,LHCb,HADES21a,PANDA21a,Xu22a}.

It is worth mentioning that
the information about 
radiative decays of baryons is very scarce.
The known decays of the decuplet baryons are 
restricted to a pair of cases:
$\Sigma^0 (1385) \to \gamma \Lambda (1116)$ and 
$\Sigma^+ (1385) \to \gamma \Sigma^+ (1193)$ decays, apart from the
$\Delta(1232) \to \gamma^\ast N$ decays~\cite{HADES21a,DecupletDecays,DecupletDalitz}.
It is then important to develop theoretical 
models which may be used to interpret the future
experiments in the timelike region
and measurements of baryon Dalitz decay rates and widths.
At the moment, there are only a few models available for 
baryon transitions in the timelike region~\cite{Sakurai60,Feinberg58,Butler93,Williams93,Shyam10,Zetenyi12,Granados17,Junker20a,Husek20a,Salone20,Zetenyi21a,Xu22a}.    

The present work is dedicated to the theoretical study of the 
$B^\prime \to \ell^+ \ell^- B$ decays, 
where $B^\prime$ is a baryon decuplet member and $B$ is a baryon octet member.
Our calculations are based on the covariant spectator 
quark model~\cite{Nucleon,NSTAR17,Omega}, 
where the contribution of the quark core 
is complemented with an $SU(3)$ contribution 
from the pion and kaon clouds, 
extending previous calculations 
where the kaon cloud was omitted~\cite{DecupletDalitz}.
The inclusion of the kaon cloud is motivated
by estimates of the radiative decays, where it was shown
that kaon cloud effects provide significant contributions 
to the $\Sigma^0 (1385) \to \gamma \Lambda (1116)$,
$\Sigma (1385) \to \gamma \Sigma (1193)$, and 
$\Xi (1530) \to \gamma \Xi (1318)$ decays~\cite{DecupletDecays}.
We start our study analyzing the Dalitz decays 
into di-electrons ($B^\prime \to e^+ e^- B$), since they 
are favored kinematically.
At the end, we make also predictions 
for the $B^\prime \to \mu^+ \mu^- B$ decays.
Our analytic expressions are compared 
with expressions derived within the $U$-spin, 
$SU(3)$ and $SU(6)$ symmetries.

Concerning the theoretical study 
of the octet baryon to the octet baryon transitions
in the spacelike region, including the photon point ($q^2=0$), 
there are studies based on a multitude of frameworks, 
including nonrelativistic
and relativistic quark models~\cite{Koniuk80,Darewych83,Kaxiras85,Sahoo95,Wagner98,Bijker00},
Dyson-Schwinger equations~\cite{Alepuz18}, 
lattice QCD simulations~\cite{Alexandrou08,Boinepalli09},
QCD sum rules~\cite{Aliev06,Wang09},
Skyrme and soliton models~\cite{Schat95,Kim20}, 
$SU(3)$ flavor models~\cite{Xu22a,Keller12,Wang21a},
chiral perturbation theory   
and large-$N_c$ models~\cite{Butler93a,Holmberg18,Lebed11}.
In general estimates solely on quark models 
underestimate the data. The estimates get closer to 
the data when the meson cloud excitations of the bare cores
are taken into account~\cite{DecupletDecays,DecupletDalitz}.

The covariant spectator quark model has been used
in the study of the electromagnetic structure 
of nucleon resonances ($\gamma^\ast N \to N^\ast$),
as well as elastic ($\gamma^\ast B \to B$) and inelastic 
($\gamma^\ast B \to B^\prime$) transitions between the baryon states.
The described nucleon resonances include the states $\Delta(1232)$,
$N(1440)$, $N(1520)$, $N(1535)$,
$\Delta(1600)$, $\Delta(1620)$, $N(1650)$, $N(1700)$ and $\Delta(1700)$, 
among others~\cite{NSTAR17,NDelta,NDeltaD,LatticeD,Spacelike12}.
As for hyperons, the model has been applied to the 
octet baryons, the decuplet baryons, transitions
between the octet and decuplet baryons, the
decays of some hyperons~\cite{Omega,Octet2Decuplet,Octet1,Octet2,LambdaSigma0,Jido12,DeltaDFF}
and $e^+ e^- \to B \bar B$ reactions~\cite{HyperonFF,Omega3FF}.
The model has also been used in the studies of nucleon deep inelastic
scattering~\cite{Nucleon,DIS2}
and of the electromagnetic
and axial structure of baryons in vacuum and in a nuclear medium~\cite{Octet2,Axial1}.
The framework provides an alternative to valence quark models 
because it takes into account also meson cloud excitations 
of the baryon cores~\cite{Octet2Decuplet,DecupletDecays,Octet1,Octet2}.
The model for the octet baryon ($B$)
to decuplet baryon ($B'$) transition~\cite{Octet2Decuplet,DecupletDecays} 
is here extended to the timelike region following 
Ref.~\cite{DecupletDalitz}, with minor modifications.
The $B' \to \gamma^\ast B$ transitions can be regarded 
a modification of the $\Delta(1232) \to \gamma^\ast N$ transition,
by the quark flavor content of the baryon states.
As for the nucleon and $\Delta(1232)$ systems, we assume 
that the octet and the decuplet baryon states are dominated by
the $S$-state quark-diquark components,
as will be discussed in the following sections.
The valence and meson cloud contributions 
to the transition form factors are analytically extended to
the timelike region, using the 
magnetic dipole form factor in terms of $q^2$ 
and the $\gamma^\ast B$ invariant mass $W$.
The results are then used to determine the 
Dalitz decay rates and Dalitz decay widths 
associated with all the decuplet baryon states.
The dependence in terms of running mass $W$, 
which may differ from the mass of the pole ($M_{B'}$) is important,
because the measurements in the timelike region 
are performed in terms of the $\gamma^\ast B$ invariant mass,
and the determination of the Dalitz decay width
requires the integration 
of functions near the pole ($W=M_{B'}$)~\cite{HADES17}.

\begin{figure}[t]
\vspace{.4cm}
\includegraphics[width=2.8in]{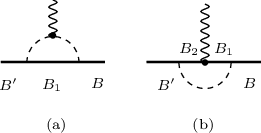}
\caption{\footnotesize
Meson cloud contributions
for the electromagnetic transition form factors.
Between the initial octet ($B$) and
final decuplet ($B'$) baryon states,
there are several possible intermediate baryon states:
$B_1$ in diagram (a); $B_1$ and $B_2$ in diagram (b).
}
\label{figMesonCloud}
\end{figure}

We conclude at the end that, the kaon cloud contributions 
increase about 30\%--40\%
the $\Sigma (1385)$ to   $\Sigma (1193)$ Dalitz decays,
and about 50\%--70\% the $\Xi (1530)$ to  $\Xi (1318)$ Dalitz decays.
We infer also that the kaon cloud effects are 
particularly important for the $\Sigma^0 (1385) \to e^+ e^- \Lambda (1116)$ decay,
because the kaon cloud contribution is enhanced in 
the diagram where the photon interacts directly with 
the intermediate baryon core [diagram (b) in Fig.~\ref{figMesonCloud}].
As a consequence, the contribution to the Dalitz decay 
width increases in about 27\%, improving significantly 
the estimates which include only pion cloud effects.
We conclude also that, measurements of the
$\Sigma^+ (1385) \to e^+ e^- \Sigma^+(1193)$, $\Sigma^{0}(1385) \to e^+ e^- \Sigma^0 (1193)$, 
$\Xi^0 (1530) \to e^+ e^- \Xi^0(1318)$ and $\Xi^- (1530)\to e^+ e^- \Xi^- (1318)$
decay widths may be expected to be measured
in a near future at HADES and PANDA~\cite{HADES21a}.
Based on our estimates of the branching ratios
and the experimental limits, 
we can also expect to observe the
$\Delta(1232) \to \mu^+ \mu^- N$ and $\Sigma^0  (1385)\to \mu^+ \mu^- \Lambda (1116)$ 
decays at BESIII and LHCb in the near future~\cite{LHCb,BESIIIreview}.

The present work is organized as follows:
in the next section, we present the formalism associated 
with the radiative and Dalitz decays of decuplet baryons
into octet baryons. 
The formalism of the covariant spectator quark model
is presented in Sec.~\ref{secCSQM}, 
while the expressions associated with the bare
valence quark and meson cloud contributions are discussed in 
Secs.~\ref{secValence} and \ref{secMesonCloud}, respectively.
In Sec.~\ref{secResultsFF}, we present our numerical 
results for the transition form factors in terms of $q^2$.
The estimates of the radiative decays, 
the Dalitz decay rates and the Dalitz decay widths 
into di-electron and di-muon pairs are presented 
and discussed in Sec.~\ref{secResults}.
The outlook and conclusions are given in Sec.~\ref{secConclusions}.
Additional information is also given in Appendixes~\ref{appQuarkFF}
to~\ref{appKaonCloud}.


\section{Radiative and Dalitz decay of decuplet baryons into octet baryons}
\label{secDalitzF}

In this section we discuss the formalism associated
with the analysis of the radiative ($B^\prime \to \gamma B$) 
and Dalitz dilepton ($B^\prime \to \ell^+ \ell^- B$) decays 
of a decuplet baryon member $B^\prime$ into an octet baryon member $B$. 
The formalism described below is valid in general for 
$\frac{3}{2}^+ \to \frac{1}{2}^+$ baryon transitions,  
using the notation $J^P$, where $J$ represent the spin and $P$ the parity.
As before, $M_{B^\prime}$ and $M_B$ represent the mass of $B^\prime$ 
and $B$, respectively.

We start with the Dalitz decay into di-electrons \mbox{($\ell =e$).}
In the following we use the indices $\gamma^\ast B$, $e^+ e^- B$
and $\gamma B$ to describe the decays 
$B^\prime \to  \gamma^\ast B$, $B^\prime \to e^+ e^- B$,  
and $B^\prime \to  \gamma B$, in that order, 
when the baryon $B^\prime$ is well identified.
Later on we discuss the Dalitz di-muon decays ($\ell =\mu$).

The Dalitz decay $B^\prime \to e^+ e^- B$ 
depends on the invariant squared four-momentum $q^2$ of the di-electron pair ($e^+ e^-$), 
and it is then limited by the $q^2 \ge 4 m_e^2 >0$,
where $m_e$ is the electron mass.  
The determination of the Dalitz decay width 
requires the use of the function $\Gamma_{\gamma^\ast B} (q,W)$,
where $W$ is the energy associated with 
the $\gamma^\ast N$ state and $q = \sqrt{q^2}$.
The Dalitz decay is interpreted as a consequence of a 
decay of a virtual timelike photon into a pair 
of electrons ($\gamma^\ast \to e^+ e^-$).     

Notice that, in the timelike region,  
we replace the dependence on the mass $M_{B'}$ by $W$,
in order to extend the discussion of the cases 
where the $\gamma^\ast B$ invariant energy 
does not correspond to the pole $M_{B'}$.

The function $\Gamma_{\gamma^\ast B} (q,W)$ 
is determined by~\cite{Dohrmann10,Timelike2,Timelike,HADES21a,Krivoruchenko01}
\ba
\Gamma_{\gamma^\ast B} (q,W)
= \frac{\alpha}{16} \frac{(W + M_B)^2}{W^3 M_B^2} 
\sqrt{y_+ y_-} y_- |G_T (q^2,W)|^2, \nonumber \\
\label{eqGammaG}
\ea
where $\alpha \simeq 1/137$ is the fine-structure constant, and
\ba
y_\pm = \sqrt{(W \pm M_B)^2 - q^2}.
\ea
The function $ |G_T (q^2,W)|^2$ takes the form
\ba
& &
|G_T (q^2,W)|^2 = \nonumber \\
& &
 |G_M (q^2,W)|^2 +  |G_E (q^2,W)|^2 + \frac{q^2}{2W^2} |G_C (q^2,W)|^2,
\nonumber \\
\label{eqGT2} 
\ea
where $G_M$, $G_E$ and $G_C$ are respectively 
the magnetic dipole, the electric quadruple and 
the Coulomb quadrupole $\gamma^\ast B \to B^\prime$ transition 
form factors, as defined by Jones and Scadron~\cite{Jones73,Devenish76}.

The Dalitz decay width $\Gamma_{e^+ e^-} (W)$ is 
determined by the integral, 
\ba
\Gamma_{e^+ e^- B}(W) = \int_{2 m_e}^{W -M_B} 
\Gamma^\prime_{e^+ e^- B} (q,W)\, dq,
\label{eqGammaInt}
\ea
where 
\ba
\Gamma^\prime_{e^+ e^- B} (q,W) 
\equiv 
\frac{d \Gamma_{e^+ e^- B}}{d q} (q,W),
\ea
represents the Dalitz decay rate function.

The function $\Gamma^\prime_{e^+ e^- B}$ can be evaluated 
using~\cite{HADES17,Dohrmann10,Timelike2,Wolf90,HADES21a}
\ba
\Gamma^\prime_{e^+ e^- B} (q,W) = \frac{2 \alpha}{3 \pi q}
\Gamma_{\gamma^\ast B} (q,W). 
\label{eqGammaP}
\ea

Notice in Eq.~(\ref{eqGammaInt}), that the integration 
in $q$ is limited to the physical region $4 m_e^2 \le q^2 \le (W-M_B)^2$,
where the momenta of the baryons $B^\prime$ and $B$ are well defined.
The upper limit corresponds to the limit where the 
photon three-momentum vanishes (${\bf q}= {\bf 0}$).
In that limit the functions $G_E$ and $G_C$ 
are correlated and $G_M$ is finite~\cite{Timelike,Siegert1,Siegert2,Siegert4,Jones73}.

The radiative decay width is defined by the function $\Gamma_{\gamma^\ast B}$, 
from Eq.~(\ref{eqGammaG}), when $q^2=0$~\cite{Dohrmann10,Wolf90},
\ba
\Gamma_{\gamma B} (W) \equiv \Gamma_{\gamma^\ast B}(0,W).
\label{eqGamma0}
\ea
The baryon $B'$ radiative decay ($B' \to \gamma \, B$)
measured in the experiments,  corresponds to
the result from Eq.~(\ref{eqGammaG})
in the limits $W= M_{B'}$ and $q^2=0$, 
\ba
\Gamma_{\gamma B} \equiv \Gamma_{\gamma B} (M_{B'}) = 
\Gamma_{\gamma^\ast B} (0, M_{B'}).
\label{eqGammaPole}
\ea

From the formalism discussed above, we conclude that the 
key ingredient to the calculation of the Dalitz decay widths
and the radiative decay widths is the determination of the transition 
form factors in the regime $q^2 \ge 0$.
Once one has a model for the transition 
form factors for $q^2 \ge 0$ in terms of a running mass $W$,   
we can calculate $|G_T (q^2,W)|$, and consequently $\Gamma_{\gamma B} (W)$, 
the Dalitz decay rate $\Gamma^\prime_{e^+ e^- B}(q,W)$, 
and the Dalitz decay width $\Gamma_{e^+ e^- B}(W)$.

An important aspect of the octet baryon to decuplet baryon electromagnetic 
transitions is that the magnetic dipole form factor is the dominant
contribution to $|G_T (q^2,W)|$.
Taking the $\gamma^\ast N \to \Delta(1232)$ transition as an example, 
it is known that near $q^2=0$, $G_E$ is about 
3--5\% of $G_M$~\cite{Siegert1,Siegert2}. 
In these conditions $G_E$ can be neglected in a first approximation.
As for $G_C$, the magnitude near $q^2=0$ is larger
and can be estimated using Siegert's theorem as 
$G_C = \frac{2 M_\Delta}{M_\Delta -M_N} G_E \simeq 
(0.03$---$0.05) \frac{2 M_\Delta}{M_\Delta -M_N} G_M$, or 
25-- 42\% of $G_M$~\cite{Siegert1}. 
Notice, however, that the $G_C$ contribution to $|G_T|^2$  
is suppressed kinematically by the factor $q^2/(2M_\Delta^2)$.
An estimate based on the relation between $G_E$ and $G_C$
near $q^2= (M_\Delta -M_N)^2 \simeq 0.086$ GeV$^2$,
indicates that the term in $|G_C|^2$ contribute to  $|G_T|^2$
with about $2 |G_E|^2$.\footnote{According 
to Siegert's theorem, $G_E = \frac{M_\Delta -M_N}{2M_\Delta} G_C$ 
near the pseudothreshold $q^2= (M_\Delta -M_N)^2$~\cite{Siegert4,Siegert1,Siegert2}.
Since this point is very close to $q^2=0$, 
we can use the relation to estimate $\frac{q^2}{2 W^2} |G_C|^2$, 
when $W$ is close to $M_\Delta$.
We obtain then $\frac{q^2}{2 W^2} |G_C|^2 = \frac{(W-M_N)^2}{2 W^2}  |G_C|^2
\simeq 2 |G_E|^2$.}
The combination of the quadrupole form factors 
to $|G_T|^2$ can then be estimated as $3 |G_E|^2$
in the region between $q^2=0$ and $(M_\Delta -M_N)^2$.
From the ratio $|G_E|/|G_M| \simeq 0.03$---0.05, 
we can conclude that $|G_T|^2 \simeq |G_M|^2$,
apart the corrections of 0.3--0.8\%.
The conclusion is then that in the Dalitz decay,  
the dominance of the magnetic dipole form factor 
is a good approximation.

Using the dominance of magnetic dipole form factor 
($G_E = G_C \simeq 0$) we can replace 
in Eq.~(\ref{eqGammaG}), $|G_T|^2$ by $|G_M|^2$.
The radiative decay width (\ref{eqGammaPole}) 
can also be reduced to 
\ba
\Gamma_{\gamma B} =
\frac{\alpha}{16} 
\frac{(M_{B'}^2 -M_B^2)^3}{M_{B'}^3 M_B^2} |G_M (0, M_{B^\prime})|^2.
\label{eqGammaB}
\ea

For the description of the Dalitz decay into 
a pair of muons, we consider the expression~\cite{Krivoruchenko01} 
\ba
\Gamma^\prime_{\mu^+ \mu^- B} (q,W) = \frac{2 \alpha}{3 \pi q} C_\mu (q^2)
\Gamma_{\gamma^\ast B} (q,W), 
\label{eqGammaP2}
\ea
where
\ba
C_\mu (q^2) = \left(1 + \frac{2 \mu^2}{q^2} \right)
\sqrt{1 - \frac{4 \mu^2}{q^2}}, 
\label{eqCmu}
\ea
and $\mu$ is mass of the muon.
A similar factor, with $\mu$ replaced by $m_e$ 
should also be included in the case of the di-electron decay 
in Eq.~(\ref{eqGammaP}).
In that case, however, one can use $C_e \simeq 1$, since 
the electron mass is negligible in the comparison with 
the other scales.

The Dalitz decay width $\Gamma_{\mu^+ \mu^- B} (W)$ is calculated similarly to 
 $\Gamma_{e^+ e^- B} (W)$ from Eq.~(\ref{eqGammaInt})
\ba
\Gamma_{\mu^+ \mu^- B}(W) = \int_{2 \mu}^{W -M_B} 
\Gamma^\prime_{\mu^+ \mu^- B} (q,W) dq,
\label{eqGammaInt2}
\ea 
where the limits of integration are redefined according to 
the value of the muon mass.

From the reduction of the integration interval,
from $[2 m_e, W-M_B]$ to $[2 \mu, W-M_B]$,  
and the weight function $C_\mu \le 1$, one can anticipate 
that the di-muon Dalitz decay width is smaller than 
the di-electron decay width.
Notice also that the function $C_\mu$ reduces strongly
the function  $\Gamma^\prime_{\mu^+ \mu^- B} (q,W)$
in the threshold of the integration ($q= 2 \mu$).


\section{Covariant spectator quark model}
\label{secCSQM}

We discuss now the formalism used in 
the present work in the calculations 
of the $\gamma^\ast B \to B^\prime$ transition form factors, 
based on the covariant spectator quark model.
The details are discussed in the next two sections.

The covariant spectator quark model is a 
constituent quark model derived from 
the covariant spectator theory~\cite{Nucleon,Gross,NSTAR17}.
Within the framework, the baryons are described 
as three-constituent quark systems, where a quark 
is free to interact with electromagnetic fields
in impulse approximation.
The electromagnetic interaction with the quarks
is described in terms of quark form factors,
that simulate the structure associated with
the gluon and quark-antiquark dressing of the quarks.
The constituent quark form factors are parametrized 
using a vector meson dominance (VMD) structure 
calibrated in the study of the electromagnetic structure 
of the nucleon and baryon decuplet~\cite{Nucleon,Omega}.

When we consider the electromagnetic interaction
in the  covariant spectator theory, 
we can integrate over the quark-pair degrees
of freedom, since they are spectators in the interaction,
and reduce the system to a diquark-quark state,
where the diquark can be represented 
as an on-mass-shell particle with effective mass
$m_D$~\cite{Nucleon,Nucleon2}. 
The baryon wave functions are derived from 
spin-flavor-radial symmetries associated with the diquark-quark configurations.
The baryon states with quantum numbers $\frac{1}{2}^+$ and  $\frac{3}{2}^+$
can be represented as combinations of $S$ 
and $D$ states~\cite{Nucleon,Nucleon2,NDelta,NDeltaD,DeltaDFF}.
The radial wave functions are 
determined phenomenologically
by experimental data, or lattice QCD results for some 
ground state systems~\cite{NSTAR17,Nucleon,Omega,LatticeD,NDeltaD,Octet1}.

The  $\Delta(1232) \to \gamma^\ast N$ transition in particular, 
can be interpreted, 
as in nonrelativistic quark models,
as dominated by the flip of the spin of a quark 
from the initial state (spin 1/2) to the final state (spin 3/2),
which can justify the dominance of the transition
by the magnetic dipole form factor
$G_M$~\cite{Aznauryan12,NSTAR,Burkert04,NSTAR17,NDelta,NDeltaD,LatticeD,Pascalutsa07}.
Small corrections, including angular momentum
excitations can induce small contributions 
to the electric ($G_E$) quadrupole form factor
and to $\frac{|{\bf q}|}{2M_\Delta} G_C$, where $G_C$ 
is the Coulomb  quadrupole form factor.
The non-zero results for these functions 
are interpreted as a manifestation 
of the $\Delta(1232)$ 
deformation~\cite{Becchi65,Glashow79,Isgur82,Krivoruchenko91,NDeltaD,Tiator01,Alexandrou08,DeltaDFF,Pascalutsa07}.
We notice, however, as discussed in the previous section that, 
the measured magnitudes of the 
electric ($G_E$) and Coulomb ($G_C$) quadrupole form factors
provide only small contributions  
to the function $G_T (q^2,M_\Delta)$ near $q^2=0$, 
and that in practice we can 
use the approximation $|G_T|\simeq |G_M|$. 
We conclude then, as discussed extensively in the literature, 
that the $\Delta(1232) \to \gamma^\ast N$ transition 
is dominated by the magnetic dipole form 
factor~\cite{Aznauryan12,Burkert04,Drechsel07,NDeltaD,LatticeD,Siegert1}.

It is important also to mention that the calculations 
based on the three-quark wave functions are not the 
full story, and that other effects have to be taken 
into account if we want to describe
$\gamma^\ast B \to B^\prime$  transition 
for arbitrary baryon states, $B$ and $B^\prime$.
Even though our quarks have structure, including meson cloud 
or quark-antiquark dressing of the quarks, 
there are processes involving the meson cloud dressing that are not taken 
explicitly into account.
The processes in which there is a meson exchange between the different
quarks inside the baryons cannot be represented by the quark dressing due to the meson cloud. 
These processes are regarded in our formalism,
as a meson emitted by a baryon state and absorbed by another baryon state,
based on a baryon-meson molecular 
picture~\cite{Octet2Decuplet,DecupletDecays,Timelike2,NSTAR17}.
The importance of the effect of the meson cloud dressing has been 
discussed in different frameworks and emphasizes the importance of the 
interplay between quark models and dynamical 
baryon-meson reaction models, where the baryon-meson interaction 
is taken into account explicitly~\cite{Burkert04,JDiaz07,Tiator01}.

The corollary of the previous discussion 
is that in the calculation of transition form factors, 
one needs to take into account 
the contributions of the valence quark core 
and the contributions of the meson cloud,  
which surrounds the baryon cores.
Considering now the octet baryon to decuplet baryon 
electromagnetic transitions ($\gamma^\ast B \to B^\prime$), we assume that
similarly to the $\gamma^\ast N \to \Delta(1232)$ transition,
those transitions are dominated by the magnetic dipole form factors,
since the $S$-wave components of the wave functions are also dominant.
In these conditions, we can use the 
decomposition~\cite{PPNP2023,DecupletDalitz,DecupletDecays,NDelta,NDeltaD,Timelike2},  
\ba
G_M (q^2,W) = G_M^{\rm B} (q^2,W) + G_M^{\rm MC} (q^2,W), 
\label{eqGM-total}
\ea
where B labels the contribution of the valence quark core 
(bare contribution) and MC labels the contribution 
associated with the meson cloud dressing.

The expressions for $G_M^{\rm B} (q^2,W)$ and $G_M^{\rm MC} (q^2,W)$
are discussed in Secs.~\ref{secValence} and~\ref{secMesonCloud}, respectively.
In Eq.~(\ref{eqGM-total}), $G_M^{\rm B} (q^2,W)$ and $G_M^{\rm MC} (q^2,W)$ 
have different falloffs for very large $|q^2|$.
The dominant valence quark contribution follows 
$G_M^{\rm B} (q^2,W) \propto 1/q^4$, 
as expected from a perturbative QCD (pQCD) 
analysis~\cite{Brodsky,Carlson}.
The meson cloud contributions fall off more quickly,
with  $G_M^{\rm MC} (q^2,W) \propto 1/q^8$,  
as will be discussed in Sec.~\ref{secMesonCloud}.

The meson cloud contributions for the $\gamma^\ast B \to B^\prime$
transitions are dominated near $q^2=0$ by the pion 
and kaon clouds and have significant contributions to  
the radiative decay widths~\cite{DecupletDecays}.
The pion cloud contributions for the Dalitz di-electron decays 
were studied in a previous work~\cite{DecupletDalitz}.
In the present work, we extend the model 
with the explicit inclusion of the kaon cloud effects.
In the following sections, we present 
a summary of the formalism, and refer the reader
to the appendixes and to Ref.~\cite{DecupletDalitz}


\section{Valence quark contributions}
\label{secValence}

We consider now the contribution from the valence quarks 
to the $\gamma^\ast B \to B^\prime$ transition.
We use the wave functions associated with the 
decuplet baryon member $\Psi_{B^\prime} (P_+, k)$  and 
the octet baryon member  $\Psi_{B} (P_-,k)$,
where $P_+$  and $P_-$ are 
the final and initial baryon momenta, respectively, 
and $k$ is the diquark momentum. 
Recall that the diquark is on-mass-shell 
in the covariant spectator quark model.
The wave functions are expressed into diquark-quark states.
For simplicity, we suppress in the wave functions the labels associated 
with the flavor and spin. 
As mentioned already, we assume that the octet baryon and 
the decuplet baryon states are dominated by the diquark-quark $S$-wave state.
This assumption is consistent with the dominance 
of the magnetic dipole form factor observed 
in the $\gamma^\ast N \to \Delta (1232)$ transition.
The explicit expressions for $\Psi_{B^\prime} (P_+, k)$ and $\Psi_{B} (P_-,k)$,
in the $S$-wave approximation, are presented
in Ref.~\cite{DecupletDalitz}.

\subsection{Transition current}

The $\gamma^\ast B \to B^\prime$ transition current 
in the relativistic impulse approximation takes 
the form~\cite{Nucleon,Nucleon2,Omega}
\ba
J^\mu = 3 \sum_{\Gamma} \int_k \overline \Psi_{B'} (P_+,k) j_q^\mu (q) \Psi_B(P_-,k),
\label{eqJtrans}
\ea
where $j_q^\mu $ is the quark current operator, 
depending on momentum transfer $q= P_+-P_-$, 
and the wave functions are expressed in terms of
diquark-quark states~\cite{Nucleon,NDelta,NSTAR17}.
There is no ambiguity in the notation for $q$,  
because the sub-indexes $q$ are replaced by $q=u,d,s$. 
The integration symbol represents the covariant integration on 
the on-mass-shell diquark momentum ($k$),
and the sum is over the diquark polarization states,
including the scalar and vector components.
The factor 3 takes into account the sum in the quarks 
based on the wave function symmetries.

The quark current $j_q^\mu$ ($q=u,d,s$) includes the electromagnetic structure 
of the constituent quarks and can 
be expressed in the form~\cite{Nucleon,Omega}
\ba
j_q^\mu (q) = j_1 (q^2)\gamma^\mu + j_2 (q^2)\frac{i \sigma^{\mu \nu} q_\nu}{2 M_N},
\ea
where $j_i$ ($i=1,2$) are the Dirac and Pauli electromagnetic flavor
operators, respectively, and act on 
the third-quark component of the wave function.
In Eq.~(\ref{eqJtrans}) $M_N$ represents the nucleon mass.
Notice that we use here $q^2= -Q^2$, instead of $Q^2$,
since we are focusing on the timelike kinematics.

The components of the quark current $j_i$ ($i=1,2$) 
can be decomposed as the sum of operators 
\ba
j_i(q^2)=
\sfrac{1}{6} f_{i+} (q^2)\lambda_0
+  \sfrac{1}{2}f_{i-} (q^2) \lambda_3
+ \sfrac{1}{6} f_{i0} (q^2)\lambda_s,
\nonumber \\
\label{eqJq}
\ea
where
\ba
&\lambda_0=\left(\begin{array}{ccc} 1&0 &0\cr 0 & 1 & 0 \cr
0 & 0 & 0 \cr
\end{array}\right), \hspace{.3cm}
&\lambda_3=\left(\begin{array}{ccc} 1& 0 &0\cr 0 & -1 & 0 \cr
0 & 0 & 0 \cr
\end{array}\right),
\nonumber \\
&\lambda_s = \left(\begin{array}{ccc} 0&0 &0\cr 0 & 0 & 0 \cr
0 & 0 & -2 \cr
\end{array}
\right),
\label{eqL1L3}
\ea
are the flavor operators.
These operators act on the quark wave function in flavor space,
$q=  (\begin{array}{c c c} \! u \, d \, s \!\cr
\end{array} )^T$.

The functions $f_{i+}$, $f_{i-}$  
represent the quark isoscalar and isovector 
form factors, respectively, based on 
the combinations of the quarks $u$ and $d$.
The functions $f_{i0}$ represent 
the structure associated with the strange quark.  
The isoscalar, isovector and strange quark form factors 
are parametrized according to a VMD form 
and are expressed in terms of 
vector meson mass poles associated with 
the $\rho$, $\omega$ and $\phi$ depending on the type ($l =\pm,0$).
For the mass of $\rho$, we use a value determined by a fit 
to the $\pi$ electromagnetic form factors, as discussed 
in the next section.
The expressions of the quark form factors 
are valid for the spacelike and timelike regions.
In the timelike region, however, the vector meson mass poles
are corrected by finite decay widths.
The explicit form for the quark form factors is 
included in Appendix~\ref{appQuarkFF}.

\subsection{Transition form factors}

The transition form factors are  calculated 
from the transition current given by Eq.~(\ref{eqJtrans})~\cite{NDelta,Octet2Decuplet}.
When we consider wave functions for the baryons $B$ and $B^\prime$, 
based on an $S$-wave diquark-quark structure, 
we conclude that only the magnetic dipole form factor  
contributes ($G_E = G_C=0$)~\cite{NDelta,NDeltaD}.
The final expression for the valence quark contribution to $G_M$,
$G_M^{\rm B}$ depends on the ($S$-wave) radial wave functions 
of the decuplet baryon member $\psi_{B^\prime} (P_+,k)$
and the octet baryon member $\psi_{B} (P_-,k)$.
In the covariant spectator quark model formalism,
the radial wave function $\psi_B(P,k)$
of a generic baryon $B$ can be parametrized 
in terms of the variable
\ba
\chi_B = \frac{(M_B -m_B)^2 -(P-k)^2}{M_B m_D},
\label{eqChiB}
\ea
where  $M_B$ and $m_D$ are the baryon mass and diquark mass, respectively.
This parametrization is possible because both the baryon $B$ and the diquark 
are on-mass-shell~\cite{Nucleon}.

The explicit expressions for $\psi_{B^\prime}$ and $\psi_{B}$ 
are presented in Ref.~\cite{DecupletDalitz}.
The parameters associated with the 
octet and decuplet baryon wave functions 
were determined in previous works by 
fits to lattice QCD data~\cite{Omega,Octet2Decuplet}.

In the following, we replace the mass of the decuplet baryon member 
$M_{B'}$ by the $W$, according to the discussion of the previous sections.

The final expression for the valence quark 
contribution to the magnetic dipole form factor 
is~\cite{NDelta,Octet2Decuplet}
\ba
G_M^{\rm B} (q^2,W) = 
\frac{4}{3 \sqrt{3}} 
\, g_v \, {\cal I} (q^2,W),
\label{eqGMbare}
\ea 
where 
\ba
{\cal I} (q^2,W)= \int_k \psi_{B'} (P_+, k)  \psi_{B} (P_-, k),
\label{eqIntBBp}
\ea
is the overlap integral of the octet baryon and 
decuplet baryon radial wave functions 
(the $W$ dependence comes from $\psi_{B'}$), 
and 
\ba
g_v = \frac{1}{\sqrt{2}} 
\left[ \frac{2 M_B}{W  + M_B}
j_1^S (q^2) + \frac{M_B}{M_N} j_2^S (q^2) \right].
\ea 
The functions $j_i^S$ ($i=1,2$) 
represent the projection of the flavor operators 
onto the flavor components of the decuplet baryon 
and the mixed symmetric component 
of the octet baryon flavor state~\cite{Octet2Decuplet}.
The explicit expressions in terms of the quark 
form factors, determined by the  
wave functions $\Psi_{B'}(P_+,k)$ and $\Psi_{B}(P_-,k)$
are presented in the second column of Table~\ref{tableJS}.
More details can be found in  Ref.~\cite{DecupletDalitz}.

\begin{table*}[t]
\begin{tabular}{l  c  c  c  c}
\hline
\hline
    & &   $j_i^S (f_{il})$  &   $j_i^S (f_{iq})$   &  
$j_i^S (f_{i-}, \Delta_i)$  \\
\hline
\hline
$\gamma^* N \to \Delta$  && $\sqrt{2}f_{i-}$  &   
 $\frac{\sqrt{2}}{3} ( 2 f_{iu} + f_{id} )$ &
$\sqrt{2} f_{i-}$ 
\\[.5cm]    
$\gamma^* \Lambda \to \Sigma^{\ast 0}$ &&
$\sqrt{\frac{3}{2}} f_{i-}$   &   
$\frac{1}{\sqrt{6}} ( 2 f_{iu} + f_{id} )$  & 
$  \sqrt{\frac{3}{2}} f_{i-}$ 
\\[.5cm] 
$\gamma^* \Sigma^+ \to \Sigma^{\ast +}$  &&
$\frac{\sqrt{2}}{6}(f_{i+} + 3 f_{i-} + 2 f_{i0})$  \spQ & 
 $ \frac{\sqrt{2}}{3}  ( 2 f_{iu} + f_{is} )$ & 
  \spQ  $ \sqrt{2} f_{i-}  -  \frac{\sqrt{2}}{3} \Delta_i$ 
\\
$\gamma^* \Sigma^0 \to \Sigma^{\ast 0}$  &&
$ \frac{\sqrt{2}}{6}(f_{i+} + 2 f_{i0})$  \spQ  &    
 $\frac{\sqrt{2}}{6}  ( 2 f_{iu} - f_{id}  + 2 f_{is} )$ &
\spQ
$\frac{\sqrt{2}}{2} f_{i-}  -  \frac{\sqrt{2}}{3} \Delta_i$ 
\\
$\gamma^* \Sigma^- \to \Sigma^{\ast -}$ &&
$\frac{\sqrt{2}}{6}(f_{i+} - 3 f_{i-} + 2 f_{i0})$    \spQ & 
    $\frac{\sqrt{2}}{3}  ( - f_{id}  +  f_{is} )$ &
 \spQ $- \frac{\sqrt{2}}{3} \Delta_i$ 
\\[.5cm]    
$\gamma^* \Xi^0 \to \Xi^{\ast 0}$ &&
$\frac{\sqrt{2}}{6}(f_{i+} + 3 f_{i-} + 2 f_{i0})$  & 
 $\frac{\sqrt{2}}{3}  ( 2 f_{iu}  + f_{is} )$  & \spQ 
$\sqrt{2} f_{i-}  -  \frac{\sqrt{2}}{3} \Delta_i$ 
\\
$\gamma^* \Xi^- \to \Xi^{\ast -}$ &&
$\frac{\sqrt{2} }{6}(f_{i+} - 3 f_{i-} + 2 f_{i0})$  &  
  $\frac{\sqrt{2}}{3}  ( - f_{id}  +  f_{is} )$  & \spQ 
  $- \frac{\sqrt{2}}{3} \Delta_i$ 
\\[0.02in]
\hline
\hline
\end{tabular}
\caption{\footnotesize
Coefficients $j_i^S$ ($i=1,2$) used to calculate the 
valence quark contributions for the transition form factors
using different representations ($l=\pm,0$ and $q=u,d,s$).
The label $\gamma^\ast N \to \Delta$ includes the $\gamma^\ast p \to \Delta^+$
and $\gamma^\ast n \to \Delta^0$ transitions 
($p$ is the proton and $n$ is the neutron).
$\Delta_i = f_{id} - f_{is}$.
}
\label{tableJS}
\end{table*}

In Table~\ref{tableJS} and along the article, 
we use the asterisk ($^*$) 
to represent the excited states of $\Sigma$ and $\Xi$, 
members of the baryon decuplet.
The label $\gamma^\ast N \to \Delta$ includes the $\gamma^\ast p \to \Delta^+$
and $\gamma^\ast n \to \Delta^0$ transitions 
($p$ is the proton and $n$ is the neutron).

The octet baryon ($\psi_B$) and decuplet baryon ($\psi_{B'}$) 
radial wave functions, presented above, ensure that 
the valence quark contribution  
to $G_M$ defined by Eq.~(\ref{eqGMbare}) 
is proportional to $1/Q^4$ for very large $Q^2$~\cite{NDelta},
consistent with estimates from perturbative QCD (pQCD)~\cite{Carlson}.

The use of the present parametrizations for $G_M^B$ and for 
the radial wave functions
is discussed in detail in our previous work on this 
subject~\cite{DecupletDalitz}.
Here, we mention only the relevant points.
The parametrization for the radial wave functions 
is justified because those functions are calibrated by lattice QCD data
for large pion masses and therefore, there is no significant contamination
from meson cloud effects.
Due to the nature of the transitions, 
in addition to the valence quark effects,  
one needs also to consider the effects associated with the pion and kaon clouds.
The present parametrizations, including the meson cloud parametrizations
of the next section, are supported by the available data 
for the octet baryon to decuplet baryon transitions,
namely by the $\gamma^\ast N \to \Delta (1232)$ transition.
Our parametrization of the $\Delta(1232)$ radial wave function 
is consistent with lattice QCD data and with 
the EBAC/Argonne-Osaka~\cite{JDiaz07,Burkert04,Aznauryan12} estimates of 
the bare core contribution to $G_M$~\cite{Octet2Decuplet,DecupletDecays,NDelta,NDeltaD,Timelike,Timelike2}.

\subsection{$SU(3)$ structure of $G_M^B$}

The valence quark contribution to $G_M$, 
determined by Eq.~(\ref{eqGMbare}) carry the $SU(3)$ symmetry structure
through the coefficients $j_i^{S}$.
The $\gamma^\ast N \to \Delta$ and $\gamma^\ast \Lambda \to \Sigma^{\ast 0}$
transitions have a pseudovector character.
The coefficients $j_i^S$ are zero 
($\gamma^\ast \Sigma^{-} \to \Sigma^{*-}$ and $\gamma^\ast \Xi^{-} \to \Xi^{*-}$)
or functions proportional to $f_{i-}$ in the limit $f_{i+} \equiv f_{i-} \equiv f_{i0}$ 
[exact $SU(3)$ symmetry limit].
The expressions for $G_M^B$ are reduced to five non-zero functions 
which differ from the result for $\gamma^\ast N \to \Delta$ 
by a single coefficient $(1,\; 1/\sqrt{3},\; 1,\; 1/2,\; 0,\; 1,\;0)$
in the limit where all members of the octet have the mass $M_B$
and all members of the decuplet have the mass $W$.

The conclusion is then that, in the exact $SU(3)$ symmetry limit,
all non-zero transitions are determined by a single function, 
which we can choose as the result for the $\gamma^\ast N \to \Delta$ transition.
Next, we consider a weaker restriction of
the $SU(3)$ symmetry, the $U$-spin symmetry.

\subsection{$U$-spin symmetry}
\label{secUspin}

For the following discussion, it is convenient to replace 
the representation of the isoscalar, isovector and $f_{i0}$ 
quark form factors by the $u$, $d$, and $s$ quark form factors.
For the conversion, we use the relations  ($i=1,2$)~\cite{Nucleon,Omega}
\ba
+ \frac{2}{3} f_{iu} &= & \frac{1}{6} f_{i+} + \frac{1}{2} f_{i-}, 
\label{eqfiu}\\
- \frac{1}{3} f_{id} &= & \frac{1}{6} f_{i+} - \frac{1}{2} f_{i-}, 
\label{eqfid}\\
- \frac{1}{3} f_{is} &= & -\frac{1}{3} f_{i0}.
\label{eqfis}
\ea
The pre-factors $e_q=-   \frac{1}{3},  + \frac{2}{3}$ 
are related to the charges of the quarks $u$ ($+\frac{2}{3}$) and
$d$, $s$  ($- \frac{1}{3})$.

A useful relation is 
\ba
3 f_{i-} = 2 f_{iu} + f_{id}.
\ea

The expressions for $j_i^S$ in terms of the 
quark form factors $f_{iq}$ are presented
in the third column of Table~\ref{tableJS}.
 
The valence quark contributions to the transition 
form factors can also be estimated within the $U$-spin symmetry~\cite{Keller12,Carruthers66}.
In the $U$-spin symmetry one can replace a $s$ quark by a $d$ quark 
in the initial and final baryon states.
The $U$-spin symmetry is exact when we can replace the $f_{id}$ by $f_{is}$.

Using the notation from Eqs.~(\ref{eqfiu})--(\ref{eqfis}) and 
\ba
\Delta_i = f_{id} -f_{is},
\ea
we obtain the expressions for $j_i^S$ presented 
in the fourth column of Table~\ref{tableJS}.

The effect of the $U$-spin can be seen more clearly 
in the last column of Table~\ref{tableJS}.
The factor associated with $\gamma^\ast \Sigma^- \to \Sigma^{*-}$
and  $\gamma^\ast \Xi^- \to \Xi^{*-}$ vanish in the 
limit $\Delta_i = 0$ (or $f_{id} \equiv f_{is}$),
and all the other coefficients became proportional to $f_{i-}$
(quark isovector form factor).

Additional relations can now be derived 
for the transition form factors, 
when we assume that all octet baryon members have the same mass, 
and all decuplet baryon members have the same mass.
In that case, we can write:
\ba
& &
\hspace{-1.5cm}
G_M (\gamma^* \Lambda \to \Sigma^{*0}) =
\frac{\sqrt{3}}{2}  G_M(\gamma^* n \to \Delta^0), \\
& &
\hspace{-1.5cm}
G_M (\gamma^* \Sigma^+ \to \Sigma^{*+}) =
 G_M(\gamma^* p \to \Delta^+),  \\
& &
\hspace{-1.5cm}
G_M (\gamma^* \Xi^0 \to \Xi^{*0}) =
\frac{1}{2}G_M( \gamma^* \Sigma^0 \to \Sigma^{*0}), \\
& &
\hspace{-1.5cm}
G_M ( \gamma^* \Sigma^- \to \Sigma^{*-}   ) =
G_M( \gamma^* \Xi^- \to \Xi^{*-}) \equiv 0.
\ea

The novelty of the above relations
compared with most of the $U$-spin analysis is the
direct relation between the $\gamma^* \Xi^0 \to \Xi^{*0}$
and $\gamma^* \Sigma^0 \to \Sigma^{*0}$ transitions.
Using the $\gamma^* N \to \Delta$ transition as reference, 
\ba
G_M^0 \equiv G_M( \gamma^* N \to \Delta), 
\ea
we can express all the non-zero form factors as
\ba
& &
G_M( \gamma^* N \to \Delta) = G_M^0,  \\
& &
G_M (\gamma^* \Lambda \to \Sigma^{*0})
= \frac{\sqrt{3}}{2}  G_M^0, \\
& &
G_M (\gamma^* \Sigma^+ \to \Sigma^{*+}) = G_M^0, \\
& &
G_M (\gamma^* \Sigma^0 \to \Sigma^{*0}) = \frac{1}{2} G_M^0, \\
 & &
G_M (\gamma^* \Xi^0 \to \Xi^{*0}) = G_M^0. 
\ea 

The previous relations can help to understand 
the relative magnitude of the different radiative decay widths.
These relations, however, are not precise enough to estimate 
the final results, for the following three main reasons:
(i) the meson cloud effects are relevant
and change the magnitude of $G_M$; 
(ii) the amount of meson cloud varies from
decay to decay; and (iii) the coefficient
of $|G_M(0)|^2$ involved in the calculations of $\Gamma_{\gamma B}$
depends on the baryon masses.
The last effect is a consequence of the $SU(3)$ symmetry breaking.

A consequence of the mass factor is that 
$\Gamma_{\Sigma^{\ast +} \to \gamma \Sigma^+} \simeq \Gamma_{\Delta \to \gamma N}$
and 
$\Gamma_{\Sigma^{\ast 0} \to \gamma \Sigma^0} \simeq \frac{1}{4}\Gamma_{\Delta \to \gamma N}$
are not good approximations.
In contrast,
$\Gamma_{\Sigma^{\ast 0} \to \gamma \Lambda} \simeq \frac{3}{4} \Gamma_{\Delta \to \gamma N}$ and 
$\Gamma_{\Sigma^{\ast 0} \to \gamma \Sigma^0} \simeq 
\frac{1}{4}\Gamma_{\Sigma^{*+} \to \gamma \Sigma^+}$
provide good estimates, because the mass factors are similar
and the meson cloud contributions have similar proportions.

\section{Meson cloud contributions}
\label{secMesonCloud}

The pion cloud contributions to the
$\gamma^\ast B^\prime \to B$ transitions
are estimated by the $SU(3)$ extension of 
our pion cloud model for the 
$\gamma^\ast N \to \Delta(1232)$ transition~\cite{NDelta,NDeltaD,LatticeD}.

We use in particular the results of Ref.~\cite{DecupletDecays},
where the meson cloud contributions of the diagrams of Fig.~\ref{figMesonCloud} 
are determined explicitly in the limit $q^2=0$.
The calculations of the meson cloud loops 
are based on the cloudy bag model 
(CBM)~\cite{Thomas84,Theberge83,Kubodera85,Tsushima88,Yamaguchi89,Lu98}.
More calculations of $\gamma^\ast N \to \Delta(1232)$ 
transition form factors 
based on the CBM can be found in Refs.~\cite{Lu97,Kaelbermann83,Bermuth88}.

The explicit calculations use the meson-baryon couplings  
for the possible octet baryon and decuplet baryon 
intermediate states from Fig.~\ref{figMesonCloud}.
The connection with the quark microscopic properties  
between the covariant spectator quark model and the CBM  
is performed matching the Dirac and Pauli couplings.
The expressions used for the calculations 
of the meson cloud contributions are presented 
in Appendixess~\ref{appPionCloud}, \ref{appMesonCloud}, and~\ref{appKaonCloud}.
In addition to the pion, we consider also
the contributions of the $K$-, $\bar{K}$- and $\eta$-meson
cloud contributions~\cite{DecupletDecays}.
The $\eta$-meson cloud contributions prove to be very small.

We describe next how to generalize the pion and 
kaon ($K$ and $\bar{K}$) cloud contributions to finite $q^2$.

\subsection{Pion cloud contributions}
\label{secPionCloud}

The calculation of the pion cloud contributions 
to the timelike region follows the lines 
of previous works for the $\gamma^\ast N \to \Delta(1232)$ 
transition~\cite{Timelike2,DecupletDalitz} 
with a few modifications.
We use the form, 
\ba
G_M^\pi (q^2)& =&   
G_M^{\pi {\rm a}} (0) \, F_\pi (q^2) 
\left(  \frac{\Lambda_\pi^2}{\Lambda_\pi^2 - q^2} \right)^3
\nonumber \\
& &  + \, G_M^{\pi {\rm b}} (0)  \, \tilde G_D^2 (q^2),
\label{eqGMpi}
\ea
where $G_M^{\pi {\rm a}} (0)$ 
and $G_M^{\pi {\rm b}} (0)$ are the pion contributions
for the diagrams (a) and (b) at $q^2=0$, respectively,
$F_\pi (q^2)$  is the pion electromagnetic form factor, 
$\Lambda_\pi^2 = 2.30$ GeV$^2$,
and $\tilde G_D (q^2)$ is a generalization 
of the traditional dipole form factor.
The coefficients  $G_M^{\pi {\rm a}} (0)$ 
and $G_M^{\pi {\rm b}} (0)$ are presented in Table~\ref{tableMesonCloud}.
In Eq.~(\ref{eqGMpi}), we omit the dependence on $W$, 
since the coefficients  $G_M^{\pi {\rm a}} (0)$ 
and $G_M^{\pi {\rm b}} (0)$ are determined in the physical limit ($W=M_{B'}$).

\begin{table*}[t]
\begin{tabular}{l  r r r  r r r r c c}
\hline
\hline
       &  $G_M^{\pi {\rm a}} (0)$  & $G_M^{\pi {\rm b}} (0)$ &   $G_M^{\pi} (0)$ &
          $G_M^{K{\rm a}} (0)$   &  $G_M^{K{\rm b}} (0)$  &    $G_M^{K} (0)$  &
\spQ  $G_M^{\rm B} (0,M_{B'})$  & $\alpha_\pi (\%)$ & $\alpha_K (\%)$\\
\hline
\hline
$\gamma^\ast N \to \Delta$            & 0.713 & 0.610 &  1.323 & 0.0167 & 0.0367 & 0.0534 & 1.633 \spQ & 44.0 & 1.8\\ [.2cm]
$\gamma^\ast \Lambda \to \Sigma^{* 0}$ & 0.669 & 0.358 &  1.027 & 0.0670 & 0.2768 & 0.3438 & 1.683 \spQ & 36.0 & 11.3\\ [.2cm]
$\gamma^\ast \Sigma^+ \to \Sigma^{* +}$& 0.149 & 0.513 &  0.663 & 0.1527 & 0.2640 &  0.4167 & 2.094 \spQ & 20.9 & 13.1 \\
$\gamma^\ast \Sigma^0 \to \Sigma^{* 0}$ & 0.000 & 0.270 & 0.270 & 0.1019 & 0.1001 &  0.2020 & 0.969 \spQ & 18.7 & 14.0\\
$\gamma^\ast \Sigma^- \to \Sigma^{* -}$ & $-$0.149 & 0.026 &  $-$0.124 & 0.0510 & $-$0.0638 & $-$0.0128 & $-$0.156 \spQ& 42.3 &4.4  \\[.2cm]
$\gamma^\ast \Xi^0 \to \Xi^{\ast 0}$  &   0.222  &  0.086  &   0.308   &  0.1850 &   0.5126  &  0.6976 & 2.191 \spQ & 9.6 & 21.8\\
$\gamma^\ast \Xi^- \to \Xi^{\ast -}$  & $-$0.222 &  0.084  &  $-$0.138 &  0.0370 & $-$0.1070 & $-$0.0700 & $-$0.168 \spQ & 36.7 & 18.6\\
\hline
\hline
\end{tabular}
\caption{\footnotesize
Coefficients of the meson cloud contributions.
The column with $G_M^{\rm B} (0,M_{B'})$ include the bare contribution at $q^2=0$.
The last two columns indicate the fraction of pion and kaon cloud for $G_M(0,M_{B'})$.}
\label{tableMesonCloud}
\end{table*}

The explicit form of Eq.~(\ref{eqGMpi}) for $G_M^\pi$
is motivated by the fast falloff of the pion cloud 
contribution in the spacelike region,
which may be simulated by the $1/Q^8$ falloff, 
for very large $Q^2$ (pQCD regime).
The first term includes the connection to the pion form factor 
$F_\pi (q^2)$ at low $q^2$ associated with the diagram \ref{figMesonCloud} (a).
The second term provides a simple and effective 
parametrization of the contributions 
associated with the possible baryon intermediate states,
also ruled by a falloff $1/Q^8$.
Using pQCD arguments, one can conclude 
that the systems with $n$ constituents contribute to the asymptotic form: 
$G_M \propto 1/Q^{2(n-1)}$~\cite{Carlson,Brodsky}.
Thus, the valence quark contributions ($n=3$) are dominated 
by $G_M \propto 1/Q^{4}$,  and contributions 
associated with the quark-antiquark excitations ($n=5$) 
are dominated by $G_M \propto 1/Q^{8}$ for large $Q^2$.

Compared to Refs.~\cite{Timelike2,DecupletDalitz}, 
we use a tripole function instead of a dipole
because we replace the parametrization of $F_\pi (q^2)$ 
by a function which falls off as $F_\pi \propto 1/Q^2$,
as discussed next.

For the pion electromagnetic form factor 
we use a parametrization derived from  
VMD Lagrangian associated with the pion-rho couplings 
and photon-rho couplings 
with the form~\cite{Gounaris68,Barkov85,Herrmann93,Connell95,Connell97,Donges95,Benayoun93},
\ba
F_\pi (q^2) = \frac{m_{\rho}^2}{m_{\rho}^2 - q^2 - 
i m_{\rho} \Gamma_\rho (q^2)}, 
\label{eqFpi}
\ea 
where $m_{\rho}$ is the mass of the $\rho$-meson and~\cite{Connell95,Benayoun93,Jackson64} 
\ba
\hspace{-.5cm}
\Gamma_\rho (q^2) = \Gamma_\rho^0 
\frac{m_\rho}{\sqrt{q^2}}
\left(\frac{q^2 - 4 m_\pi^2}{m_{\rho}^2 -4 m_\pi^2  } \right)^{3/2}
\theta (q^2 - 4 m_\pi^2),
\label{eqGammaRho}
\ea
where $m_\pi$ is the pion mass, $\theta$ is the Heaviside function
and $\Gamma_\rho^0$ is a constant associated with   
the $\rho \to \pi \pi$ decay at the physical $\rho$ mass ($q^2= m_\rho^2$).

In the spacelike region $\Gamma_{\rho} \equiv 0$ 
and $F_\pi (q^2) \propto 1/Q^2$ for large $Q^2$.
In the timelike region, one has $\Gamma_\rho \propto q^2$
for large $q^2$, 
consequently, we obtain  $F_\pi (q^2) \propto 1/q^2$,
for large $q^2$.

In the expressions (\ref{eqFpi}) and (\ref{eqGammaRho}),  
$m_\rho$ ($\rho$-meson dressed mass) and  $\Gamma_\rho^0$
can be calculated using a parametrization of the $|F_\pi(q^2)|^2$ data.
The physical value of $m_\rho$ is not very precise 
because one obtains different estimates depending
on the experiment ($\rho^0 \to \pi^+ \pi^-$ decay, $\rho^\pm$ decays,
hadronic production etc.)~\cite{PDG22,PDG-rho}.
In the literature, it was also discussed that  
the data associated with $|F_\pi(q^2)|^2$ cannot be described accurately by
the relativistic Breit-Wigner from Eq.~(\ref{eqGammaRho})
and it requires some additional shape parameters~\cite{PDG-rho,Benayoun93,Pisut68}.
It is also worth noticing that near the peak ($q^2 \simeq m_\rho^2 \simeq 0.6$ GeV$^2$) 
the function  $|F_\pi(q^2)|^2$ has contributions 
associated with the $\omega \to \pi \pi$ decay (isoscalar contribution)
due to the $\rho$-$\omega$ mixing.
Since our purpose is to parametrize the isovector contributions, 
we ignore those effects and determine 
the values of $m_\rho$ and $\Gamma_\rho^0$ by a fit 
to the  $|F_\pi(q^2)|^2$ in the range $-1$ GeV$^2$ $< q^2 < 1$ GeV$^2$.
For larger values of $q^2$ there are contamination 
of poles associated with higher mass excitations of the $\rho$.
From the fit, we obtain the values
\ba
m_\rho =  0.771 \; \mbox{GeV},  \hspace{1cm} 
\Gamma_\rho^0 = 0.117  \; \mbox{GeV}. 
\label{eqParameters}
\ea
The results of the fit are presented in Appendix~\ref{appPionCloud}.

The parametrization  of $|F_\pi(q^2)|$ by Eqs.~(\ref{eqFpi}) and (\ref{eqGammaRho})
is important because it includes the analytic structure associated 
with the two-pion cut in an exact form.
In addition, we ensure that we have the correct falloff for 
$F_\pi(q^2)$ for large $Q^2=-q^2$.
More details about the calibration based on Eq.~(\ref{eqGMpi})
are included in Appendix~\ref{appPionCloud}.

We discuss now the function $\tilde G_D$.
Following Ref.~\cite{Timelike2}, 
the function $\tilde G_D$ is defined by
\ba
\tilde G_D (q^2) =
\frac{\Lambda_D^4}{(\Lambda_D^2 - q^2)^2 + \Lambda_D ^2 \Gamma_D^2}, 
\label{eqGD}
\ea
where $\Lambda_D^2 = 0.9$ GeV$^2$ and 
$\Gamma_D (q^2)$ is an effective width.  
For $\Gamma_D (q^2)$, we use parametrization~\cite{Timelike2,N1520TL,N1535TL} 
\ba
\Gamma_D (q^2) = 4 \Gamma_D^0 \left(  
\frac{q^2}{\Lambda_D^2 + q^2} \right)^2
\theta (q^2),
\label{eqGammaD}
\ea
where $\theta$ is the Heaviside function
and $\Gamma_D^0 = 4 \Gamma_\rho^{\rm exp}$.
Here $\Gamma_\rho^{\rm exp} \simeq 0.149$ GeV is the experimental 
value of the $\rho$ decay width.
The form (\ref{eqGammaD}) provides a smooth continuous transition between 
the spacelike and the timelike region.\footnote{Thus, 
although the regularization generates 
spurious imaginary contributions below the 
physical thresholds ($4 m_\pi^2$, $9 m_\pi^2$ 
related to vector meson poles),  
those contributions are small at low $q^2$
(proportional to $q^4$) and interfere minimally 
with the imaginary contributions associated 
with the physical poles.}
The function $\Gamma_D (q^2)$
converge to $\Gamma_D^0$ for very large $q^2$ 
and spread out the effect of the pole 
$q^2=\Lambda_D^2$~\cite{Timelike2,DecupletDalitz}.

\subsection{Kaon cloud contributions}
\label{secKaonCloudP}

The extension of the kaon cloud contribution
to the timelike region follows lines similar
to the pion cloud contribution of Eq.~(\ref{eqGMpi}).
We use then 
\ba
G_M^K (q^2)& =& 
G_M^{K {\rm a}} (0) \, F_K (q^2) 
\left(  \frac{\Lambda_{K{\rm a}}^2}{\Lambda_{K{\rm a}}^2 - q^2} \right)^3 
\nonumber \\
& & + \, G_M^{K{\rm b}} (0)  \, \tilde G_{DK}^2 (q^2), 
\label{eqGMK}
\ea
where $G_M^{K{\rm a}} (0)$ 
and $G_M^{K{\rm b}} (0)$ are the kaon contributions
for the diagrams (a) and (b), respectively,
$F_K$ is the kaon electromagnetic form factor,  
$\Lambda_{K{\rm a}}$ is a new cutoff,  
and $\tilde G_{DK}$ is a new dipole form factor.

For the kaon form factor, we use
\ba
F_K (q^2) = \frac{\Lambda_{K}^2}{\Lambda_K^2 - q^2},
\label{eqFK}
\ea
where $\Lambda_K^2$ is determined by 
the kaon square radius $r_K^2 = 0.314 \pm 0.035$ fm$^2$~\cite{PDG22}.
One obtains then $\Lambda_K^2 \simeq 0.745$ GeV$^2$.
The monopole form  (\ref{eqFK}) of $F_K$ 
is motivated by pQCD according to the expected falloff $F_K \propto 1/q^2$ 
for a $q \bar q$ system~\cite{Carlson,Brodsky}.
The inclusion of the tripole factor is also 
motivated by pQCD arguments.
One expects that the falloff of the baryon-meson system to be $1/q^8$
due to the number of constituents (three quarks plus a quark-antiquark pair)~\cite{Carlson}.


\begin{table*}[t]
\begin{tabular}{l   c c c  c}
\hline
\hline
      &  $\frac{d \log G_M^{K{\rm a}}  }{d {\bf q}^2}  (0)$  
      &  \spQ  $\Lambda_{K{\rm a}}^2$ \spQ 
      &  $\frac{d \log G_M^{K{\rm b}} }{d {\bf q}^2}  (0)$   
      &  \spQ $\Lambda_{K{\rm b}}^2$ \spQ \\[.05cm]
\hline
\hline
$\gamma^\ast N \to \Delta$            & $-$2.89  &  1.944 & $-$2.05   & 1.951 \\[.2cm]
$\gamma^\ast \Lambda \to \Sigma^{* 0}$ & $-$3.20  &  1.618 & $-$5.40   & 0.741   \\[.2cm]
$\gamma^\ast \Sigma^+ \to \Sigma^{* +}$ & $-$2.93 &  1.889 & $-$2.37       & 1.688 \\
$\gamma^\ast \Sigma^0 \to \Sigma^{* 0}$ & $-$2.99 &  1.825 & $-$2.43       & 1.645     \\ 
$\gamma^\ast \Sigma^- \to \Sigma^{* -}$ & $-$3.15 &  1.660 &  $-$2.17 & 1.841 \\[.2cm]
$\gamma^\ast \Xi^0 \to \Xi^{\ast 0}$     & $-$2.98  &  1.829 &  $-$3.37       &  1.186 \\
$\gamma^\ast \Xi^- \to \Xi^{\ast -}$     & $-$2.84  &  2.000 &  $-$2.33  &  1.715 \\
\hline
\hline
\end{tabular}
\caption{\footnotesize
Logarithmic derivatives of the kaon cloud contributions and cutoffs $\Lambda_{Ka}$ and  $\Lambda_{Kb}$.
The derivatives are in units GeV$^{-2}$.
The square cutoffs are in units GeV$^2$.  
Details of the calculations are presented in Appendix~\ref{appMesonCloud}.}
\label{tableKaonCloud}
\end{table*}

As for  $\tilde G_{DK}$, one considers a form similar to Eq.~(\ref{eqGD}), 
\ba
\tilde G_{DK} (q^2) =
\frac{\Lambda_{K{\rm b}}^4}{(\Lambda_{K{\rm b}}^2 - q^2)^2 
+ \Lambda_{K{\rm b}} ^2 \Gamma_{K{\rm b}}^2}, 
\label{eqGDK}
\ea
where $\Lambda_{K{\rm b}}$ is a new cutoff, 
and  $\Gamma_{K{\rm b}}$ is an effective decay width defined by 
Eq.~(\ref{eqGammaD}) with $\Gamma_D  \leftrightarrow \Gamma_{K{\rm b}}$
and $\Lambda_D \to \Lambda_{K{\rm b}}$.
Also here, the falloff is $1/q^8$.

To calculate the kaon cloud contribution, 
one needs to know the coefficients $G_M^{K{\rm a}}(0)$ 
and  $G_M^{K{\rm b}}(0)$ and the cutoffs 
$\Lambda_{K{\rm a}}$ and   $\Lambda_{K{\rm b}}$,
which depend on the decuplet and octet baryon states.
The coefficients $G_M^{K{\rm a}}(0)$ and  $G_M^{K{\rm b}}(0)$ 
were determined by the previous work 
about the decuplet baryon radiative decays~\cite{DecupletDecays}
and are presented in Table~\ref{tableMesonCloud}.
The determination of the cutoffs 
$\Lambda_{K{\rm a}}$ and $\Lambda_{K{\rm b}}$ 
require a microscopic calculation of the 
kaon cloud contribution for non-zero values of $q^2$.
The method used to calculate $\Lambda_{K{\rm a}}$ and $\Lambda_{K{\rm b}}$,
is explained in the next section.
Recall that in the case of the pion cloud, 
the free parameters from (\ref{eqGMpi}),
are determined directly from the comparison
with the physical data~\cite{LatticeD,Timelike2}.

In the case of the cutoffs $\Lambda_{K{\rm a}}$ 
or $\Lambda_{K{\rm b}}$, when they have a magnitude close to $(M_{B'}-M_B)$,
one needs to regularize the contributions of those poles
to the transition form factors, including an effective width, 
as in the case of Eq.~(\ref{eqGammaD}).

In the last three columns of Table~\ref{tableMesonCloud},
we include for convenience of the discussion:
the bare contribution $G_M^{\rm B}(0, M_{B'})$
to the magnetic form factor at $q^2=0$,
and the relative contributions 
of the pion ($\alpha_\pi$) and kaon ($\alpha_K$) to
the final results also at $q^2=0$.
These contributions are estimated by
$G_M^{m}(0)/[G_M^{\pi} (0)  + G_M^K (0)+ G_M^{\rm B}(0,M_{B'})]$,
where $m=\pi, K$.

\subsubsection{Determination of the kaon cloud cutoffs}

The calculation of the kaon cloud contributions 
for finite $q^2$ requires an extension of the
calculations for the case $q^2=0$ from 
Ref.~\cite{DecupletDecays} for the case $q^2\ne 0$.
For the present discussion, it is not important
if $q^2 > 0$ or $q^2 < 0$.
One can then consider the case $Q^2 = -q^2 > 0$.
For simplicity, we use $G_M^{K\ell} (Q^2)$ to represent $G_M^{K\ell} (q^2)$.

We examine then the calculation of the diagrams from Fig.~\ref{figMesonCloud}, 
$G_M^{K\ell} (Q^2)$, where $\ell = {\rm a}, {\rm b}$,
for positive  values of $Q^2$ near $Q^2=0$.
In the present work, we are interested in small values of $Q^2$, 
since the range of the calculations is restricted 
to $-(M_{B'} -M_B)^2  \le  Q^2 \le 0$ and 
\mbox{$(M_{B'} -M_B)^2 \le 0.1$} GeV$^2$ in most cases.

In the CBM the calculation of $G_M^{K\ell} (Q^2)$ 
must be performed in a specific frame.
We choose then the Breit frame, 
because it is more accurate due to 
the kinematic approximations used 
in the calculations at $Q^2=0$, 
more specifically the static approximation
(or the heavy baryon approximation).
In the static approximation, 
\ba
Q^2 = Q_{\rm st}^2 \simeq {\bf q}^2 - (M_{B'}- M_B)^2,
\ea
where the label "st" stands for {\it static approximation}. 
The relative error in the approximation 
for \mbox{$q_0 \simeq (M_{B'}- M_B)$} 
is of the order of $- \sfrac{1}{8M_B^2}(M_{B'} - M_B)^2 - \sfrac{1}{16M_B^2} Q^2$.
A small error\footnote{In the baryon $B'$ rest frame 
the relative error is $\sfrac{Q^2}{M_{B'}^2 -M_B^2}$.} 
for small values of $Q^2$.

In these conditions, we can use the expansion, 
\be
G_M^{K\ell} (Q^2) = G_M^{K\ell} (0) +
\left[ \frac{d G_M^{K\ell} }{d {\bf q}^2} (0) \right] Q^2.
\label{eqGMell}
\ee
From the above relation, we can conclude that
the first finite correction to  $G_M^{K\ell} (0)$
can be determined by the coefficient of 
the expansion in $Q^2 = {\bf q}^2 - q_0^2$,
which can be determined by the derivative on ${\bf q}^2$
for ${\bf q}^2 = q_0^2$.   
[In the static approximation 
$\frac{d \spQ \sp }{d Q^2} \equiv \frac{d \spQ \spQ}{d Q_{\rm st}^2} 
\equiv \frac{d \spQ \sp}{d {\bf q}^2} $].

The consequence of this discussion is that,
to estimate the cutoffs associated with (\ref{eqGMK}), 
we do not need to calculate $G_M^{K\ell} (Q^2)$ 
for a specific value for $Q^2$.
It is sufficient, instead, 
to calculate the derivative 
$\frac{d \;  }{d {\bf q}^2}  G_M^{K\ell} (Q_{\rm st}^2=0)$.
The numerical values for $G_M^{K\ell} (0)$ are 
already presented in Table~\ref{tableMesonCloud}.
The expressions for  $G_M^{K\ell} (0)$
and $\frac{d G_M^{K\ell}  }{d {\bf q}^2}   (0)$
are presented in Appendix~\ref{appKaonCloud}.


For the following discussion, it is convenient to 
normalize Eq.~(\ref{eqGMell}) by the value of $G_M^{K \ell}(0)$,
\ba
\frac{G_M^{K\ell} (Q^2)}{G_M^{K\ell} (0)} 
= 1 +  \left[
\frac{d  \log G_M^{K\ell} }{d {\bf q}^2} (0) \right] Q^2.
\label{eqGMell2}
\ea 
Notice that 
$\frac{1}{F(0)}  \frac{d F}{d Q^2} (0) \equiv  \frac{d \log F}{d Q^2}(0) $ 
for a generic function $F(Q^2)$.
One concludes, then, that  what we need for our calculations 
is in fact the  logarithmic derivative.

The numerical values of the logarithmic derivatives 
of  $G_M^{K\ell}$ at $Q^2=0$
are presented in Table~\ref{tableKaonCloud}.
The value of $G_M^{K\ell} (0)$ is the same as in Table~\ref{tableMesonCloud}
and in previous works, where the meson cloud 
contributions from the CBM are calibrated by
pion cloud contributions to the $\gamma^\ast N \to \Delta$ transitions.
In Appendix~\ref{appKaonCloud}, this procedure 
is explained in detail.

To determine the unknown cutoffs from (\ref{eqGMK}),
one considers low-$Q^2$ expansions of 
the multipoles associated with the diagrams (a) and 
(b) from Fig.~\ref{figMesonCloud}.

For the diagram (a), we use
\ba
F_K (q^2) \left(  \frac{\Lambda_{K{\rm a}}^2}{\Lambda_{K{\rm a}}^2 - q^2} \right)^3 
& \simeq & 
 \left( 1 - \frac{Q^2}{\Lambda_K^2}   \right) 
\left( 1 -  3\frac{Q^2}{\Lambda_{K{\rm a}}^2}     \right) \nonumber \\
&  \simeq & 1 - \left( \frac{1}{\Lambda_K^2} + \frac{3}{\Lambda_{K{\rm a}}^2} 
\right) Q^2,  \nonumber \\
& &
\label{eqGMKa1}
\ea
drooping terms of the order $Q^4$.
From the comparison between Eqs.~(\ref{eqGMell2})
and~(\ref{eqGMKa1}), one concludes that
\ba
\frac{1}{\Lambda_K^2} + \frac{3}{\Lambda_{K{\rm a}}^2}  
\simeq  - \frac{d \log G_M^{K{\rm a}}  }{d {\bf q}^2}(0).
\ea
This relation can be used to calculate $\Lambda_{K{\rm a}}^2$ 
based on the value of $\Lambda_K^2$ 
extracted from the
experimental value for $r_K^2$, 
and the values of the logarithm derivatives from Table~\ref{tableKaonCloud}.

As for the diagram (b), one obtains
in the regime \mbox{$Q^2> 0$},  
\ba
\tilde G_{DK} (q^2) &=&  
\left( \frac{\Lambda_{K{\rm b}}^2}{\Lambda_{K{\rm b}}^2 + Q^2} \right)^4
\nonumber \\
& \simeq & 
1 - 4 \frac{Q^2}{ \Lambda_{K{\rm b}}^2 }.
\ea
From the comparison with (\ref{eqGMell2}), we conclude that
\ba
\frac{4}{ \Lambda_{K{\rm b}}^2}  = -
\frac{d \log G_M^{K{\rm b}}  }{d {\bf q}^2} (0),
\ea
which can be used to calculate $\Lambda_{K{\rm b}}^2$ 
based on the values of the logarithm derivatives 
from Table~\ref{tableKaonCloud}.

\subsubsection{Discussion about the kaon cloud contributions}

The kaon cloud contribution to the $B' \to \gamma^\ast B$ transition
can now be determined using Eq.~(\ref{eqGMK}),  
the coefficients $G_M^{K{\rm a}} (0)$ and  $G_M^{K{\rm b}} (0)$ 
from Table~\ref{tableMesonCloud}, and the square cutoffs $\Lambda_{K{\rm a}}^2$ 
and $\Lambda_{K{\rm b}}^2$ from Table~\ref{tableKaonCloud}.

The values presented in Table~\ref{tableMesonCloud} 
are based on the calculations from Ref.~\cite{DecupletDecays}.
In the case of the values associated with the kaon cloud,  
there are minor differences compared with the previous calculation.
That happens, because we recalculate $G_M^{K\ell} (0)$ using 
a more precise numerical integration algorithm, 
and because we present the results with more digits.
This is justified since the focus is on the 
kaon cloud and it is necessary to include more precision of the calculations.
In the limit $q^2=0$, the results are qualitatively equivalent
to the results from Ref.~\cite{DecupletDecays}.
The differences between the calculations from Ref.~\cite{DecupletDecays}
are apparent, namely, only in the third digit of the final results.

A comment about the results for the $\gamma^\ast \Sigma^0 \to \Sigma^{\ast 0}$ is in order.
In the Table~III from Ref.~\cite{DecupletDecays}, there was a typo in the value 
of $G_M^{K{\rm b}}(0)$ (0.010),
which is corrected in the present work to 0.1001
in Table~\ref{tableMesonCloud}.
The typo, however, did not affect the final results.
In the present work, we conclude that the  contribution 
of the kaon cloud to the transition form factor is in fact 0.2020.

We can use the results $G_M^\pi (0)$ and $G_M^K (0)$ from  Table~\ref{tableMesonCloud} 
to discuss the magnitude of the pion and kaon cloud contributions
to the transition form factors at the photon point
according to the relative contributions of the pion
($\alpha_\pi$) and the kaon ($\alpha_K$) displayed 
in the last two columns (in a percentage).
In a first analysis, we exclude the $\gamma^\ast \Sigma^- \to \Sigma^{\ast -}$ and 
$\gamma^\ast \Xi^- \to \Xi^{\ast -}$ transitions 
because the contributions of the valence quarks are very small 
as discussed in Sec.~\ref{secCSQM} 
[a consequence of $U$-spin and $SU(3)$ symmetries].

From the results from  Table~\ref{tableMesonCloud}, 
we can conclude that the pion cloud is very important 
for the $\gamma^\ast N \to \Delta$, $\gamma^\ast \Lambda \to \Sigma^{\ast 0}$ and 
$\gamma^\ast  \Sigma \to \Sigma^{\ast}$ transitions,  
with contributions of about 44\%, 36\% and 20\%, in that order.
As for the kaon cloud, it is important for the 
$\gamma^\ast \Lambda \to \Sigma^{\ast 0}$ and $\gamma^\ast \Sigma \to \Sigma^{\ast}$
transitions (10-15\%), 
and much more relevant for the $\gamma^\ast \Xi^0 \to \Xi^{\ast 0}$ transition (22\%).
These results support the need of incorporating the kaon cloud 
contributions in the calculations of transition from factors.

Concerning the $\gamma^\ast \Sigma^- \to \Sigma^{\ast -}$ and $\gamma^\ast \Xi^- \to \Xi^{\ast -}$
transitions, the meson cloud (pion or kaon) effects have 
an important relative contribution, close to 50\%, because their magnitudes 
are comparable with the valence quark contributions.


\section{Results for the transition form factors}
\label{secResultsFF}

\begin{figure*}[t]
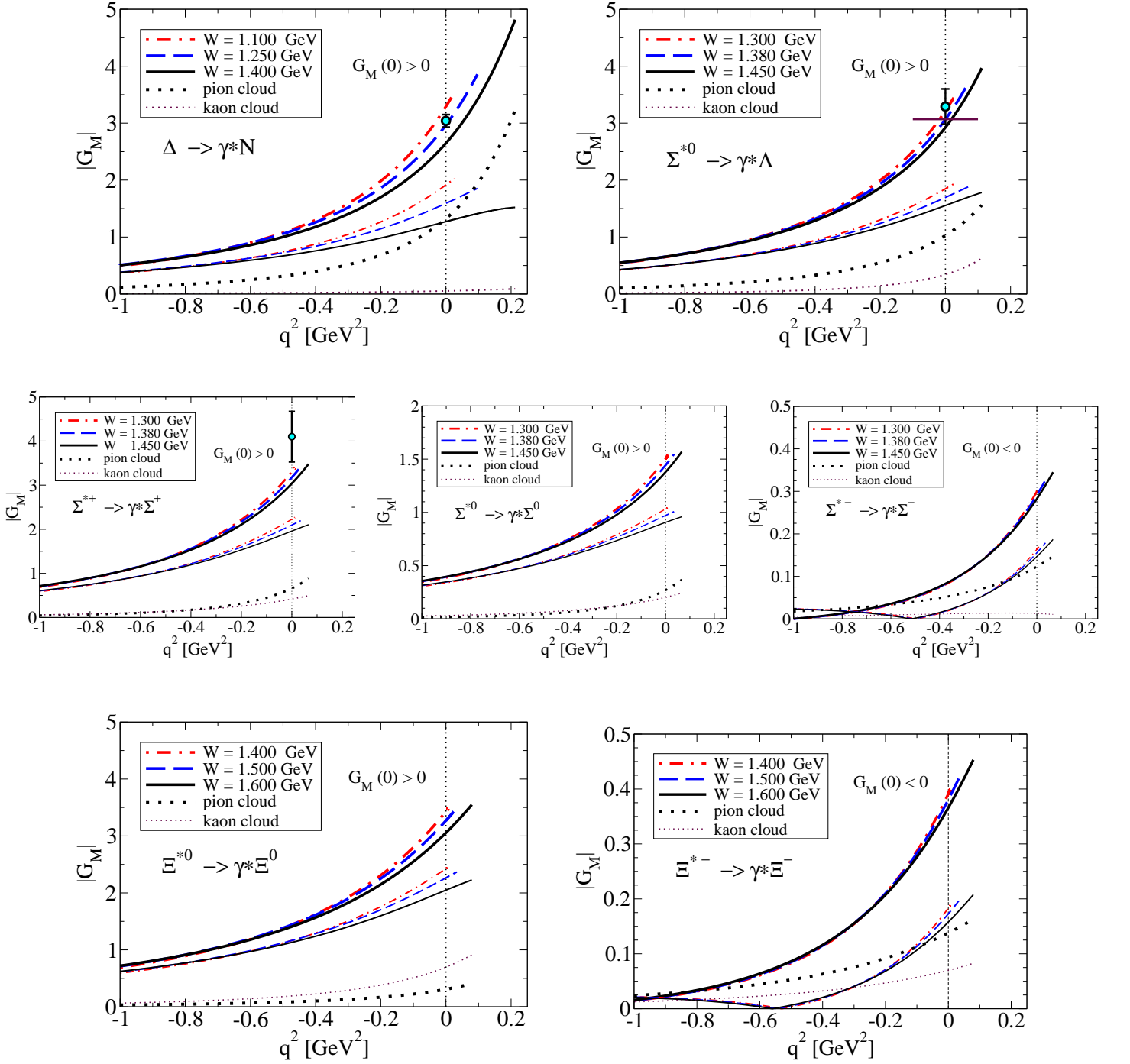

\vspace{0.5cm}
\centerline{
\mbox{
\includegraphics[width=3.1in]{GM-Nucleon-v1} \hspace{.6cm}
\includegraphics[width=3.1in]{GM-Lambda-v1} }}
\vspace{.7cm}
\centerline{
\mbox{
\includegraphics[width=2.4in]{GM-SigmaP-v1} \hspace{.15cm}
\includegraphics[width=2.4in]{GM-Sigma0-v1} \hspace{.15cm}
\includegraphics[width=2.4in]{GM-SigmaM-v1} 
}}
\vspace{.9cm}
\centerline{
\mbox{
\includegraphics[width=3.1in]{GM-Xi0-v1} \hspace{.6cm}
\includegraphics[width=3.1in]{GM-XiM-v1} }}
\caption{\footnotesize{
Magnitude of transition form factor $G_M$.
The thick lines represent the total
(valence plus pion and kaon clouds) and 
the thin lines represent the valence quark contribution.
The thick dotted line and the thin dotted 
represent the pion and kaon cloud contributions,
respectively.
The Data are from Table~\ref{tableGM0}.
}}
\label{figGM2}
\end{figure*}

In this section we present our results for 
the transition form factors between 
octet baryons and decuplet baryons.

For the numerical calculation, we use the experimental
octet baryon mass values: $M_N =0.939$ GeV,
$M_\Lambda =1.116$ GeV, $M_\Sigma = 1.192$ GeV, and 
$M_\Xi = 1.318$ GeV.
As before, $W$ represents the decuplet baryon masses.
In the calculations associated with the physical decuplet baryons,
we use the physical masses: $M_\Delta = 1.232$ GeV, $M_{\Sigma^\ast} =1.385$ GeV 
and $M_{\Xi^\ast} =1.533$ GeV.

The results are presented in Fig.~\ref{figGM2} 
for different values of $W$ close to the physical pole mass value $M_{B'}$.
The final results (bare plus meson cloud) are 
represented by the thick lines.
The bare contributions are represented by the corresponding thinner lines.
In the graphs, we include also the contributions
of the pion and kaon clouds associated with the maximum value of $W$
(dotted lines).
The pion and kaon cloud contributions 
for the first and second values of $W$ 
are described by the same line,
except that they are limited to $q^2 \le (W-M_B)^2$.
Recall that our parametrization of the meson cloud 
contributions is independent of $W$.


\begin{table*}[t] 
\begin{center}
\begin{tabular}{l  cc  cc  cc  cc  cc cc cc}
\hline
\hline
     && 
\sp\sp$\left. G_M(0)\right|_{{\rm no} K} $ 
&&  \sp\sp $\left. G_M(0)\right|_K $
 &&   \sp\sp$G_M(0)$ 
 && $|G_M(0)|_{\rm exp}$  &&
$\Gamma_{\gamma B} ({\rm keV}) $ && $\Gamma_{\gamma B}^{\rm exp} ({\rm keV})$
\\ [.05cm]
\hline
\hline
$\Delta \to \gamma N$ && 2.96 && 0.053  &&  3.01    && 
\sp\sp$3.04\pm0.11$ \cite{PDG10}
&&  644 && $660\pm47$ \cite{PDG10}  \\[.15cm]
$\Sigma^{\ast 0} \to \gamma \Lambda$ &&  2.71 && 0.344 &&   3.05   &&
\sp\sp$3.35\pm0.57$ \cite{PDG10} &&  392 && $470\pm 160$ \cite{PDG10} \\
  &&    &&  &&  &&
\sp\sp$3.26\pm0.37$ \cite{Keller12} &&  && $445\pm 102$
\cite{Keller12} \\[.15cm]
$\Sigma^{\ast +} \to \gamma \Sigma^+ $ && 2.76 && 0.417 &&  3.17 && 
\sp\sp$4.10\pm0.57$
\cite{Keller11a}  &&  149 && $250\pm70$ \cite{Keller11a} \\
$\Sigma^{\ast 0} \to \gamma \Sigma^0 $ &&  1.24 && 0.202 &&   1.44  && $< 11$~\cite{Colas75} 
&&   31 && $< 1750$~\cite{Colas75}\\
$\Sigma^{\ast -} \to \gamma \Sigma^-$ &&  $-0.280$ \sp\sp && $-0.013$\sp\sp &&
$-0.291$ \sp &&   $< 0.8$ \cite{Molchanov04} &&  1.3 && $< 9.5$  \cite{Molchanov04} 
\\ [.15cm]
$\Xi^{\ast 0} \to  \gamma \Xi^0 $ && 2.50 && 0.698 &&  3.20   &&       &&  172 &&     \\   
$\Xi^{\ast -} \to  \gamma \Xi^- $ &&  $-0.306$ \sp\sp  && $-0.070$\sp\sp && 
$-0.376$ \sp  &&  $< 4.2$ \cite{Ablikim19} &&  2.4 &&   $< 366$ \cite{Ablikim19}\\
\hline
\hline
\end{tabular}
\end{center}
\caption{\footnotesize
Results for $G_M(0)$ corresponding to the $B' \to \gamma \, B$ decays. 
$\left. G_M(0)\right|_{{\rm no} K} $ gives the result
without the kaon cloud (bare plus pion cloud).
$\left. G_M(0)\right|_K $ represents exclusively
the kaon cloud contribution (no bare core term).
The values for $|G_M(0)|_{\rm exp}$ are estimated 
using the experimental values of
$\Gamma_{\gamma B}$ ($\Gamma_{\gamma B}^{\rm exp}$).
The $\eta$ cloud contributions are in general 
negligible---the exception is the $\Xi^{\ast 0} \to  \gamma \Xi^0$ 
transition (0.086)~\cite{DecupletDecays}.
For the $\Sigma^{*0} \to \gamma \Lambda$, we consider the average $|G_M(0)|=3.29\pm0.31$
and $\Gamma_{\gamma B} = 452 \pm 86$ keV.}
\label{tableGM0}
\end{table*}

In Fig.~\ref{figGM2}, one can see 
the significant contributions from the kaon cloud
in all the transitions, except for 
the $\Delta  \to  \gamma^\ast N$ and 
$\Sigma^{\ast -} \to \gamma^\ast \Sigma^-$ transitions.
In the first case the kaon cloud effects are in 
fact very small (from both diagrams).
In the second case, there is a strong cancellation
between the contributions from the diagrams (a) and (b),
as can be confirmed by the values of 
the coefficients $G_M^{K{\rm a}}(0)$ and $G_M^{K{\rm b}}(0)$
in Table~\ref{tableMesonCloud}.

The enhancement of $|G_M|$ by the kaon cloud in the timelike region 
is more pronounced for the $\Sigma^{\ast 0} \to \gamma^\ast \Lambda$
and $\Xi^{\ast 0} \to \gamma^\ast \Xi^0$ transitions.
In the first case  the reason is due to the small magnitude
of the cutoff $\Lambda_{K{\rm b}}$  
[sharper contribution from the diagram (b) near $Q^2=0$].
In the second case, the kaon cloud coefficients are large
(see Table~\ref{tableMesonCloud}).

We do not make a detailed comparison
of our calculations with those in the literature, because the comparison was already
presented in previous works for the case 
$q^2=0$~\cite{DecupletDecays}
and for $q^2 > 0$ with the meson cloud  
restricted to the pion cloud
(see Sec.~III.E in Ref.~\cite{DecupletDalitz}).
One can, nevertheless, compare our bare estimates 
with other calculations of transition form factors for finite $q^2$ 
based on quark degrees of freedom 
as in Refs.~\cite{Alepuz18,Kim20}.

The calculations from Ref.~\cite{Alepuz18} 
are based on the Dyson-Schwinger framework.
Their results are similar to our 
estimates for $G_M^B$ for $q^2 < -$ 1 GeV$^2$.
There are also calculations based on an $SU(3)$ chiral quark-soliton 
model~\cite{Kim20}, which include some pion cloud effects.
The estimates of the contribution of the bare core for $q^2=0$
are similar to our estimates.
The results for $G_M$ including meson cloud, however,
have slower falloffs with $Q^2=- q^2$, 
when compared to our estimates.

\section{Results for the radiative and Dalitz decay widths}
\label{secResults}

In this section, we discuss our results for the radiative and
the Dalitz decay widths 
associated with the baryon decuplet $B^\prime$ decays to baryon octet $B$.


\begin{figure*}[t]
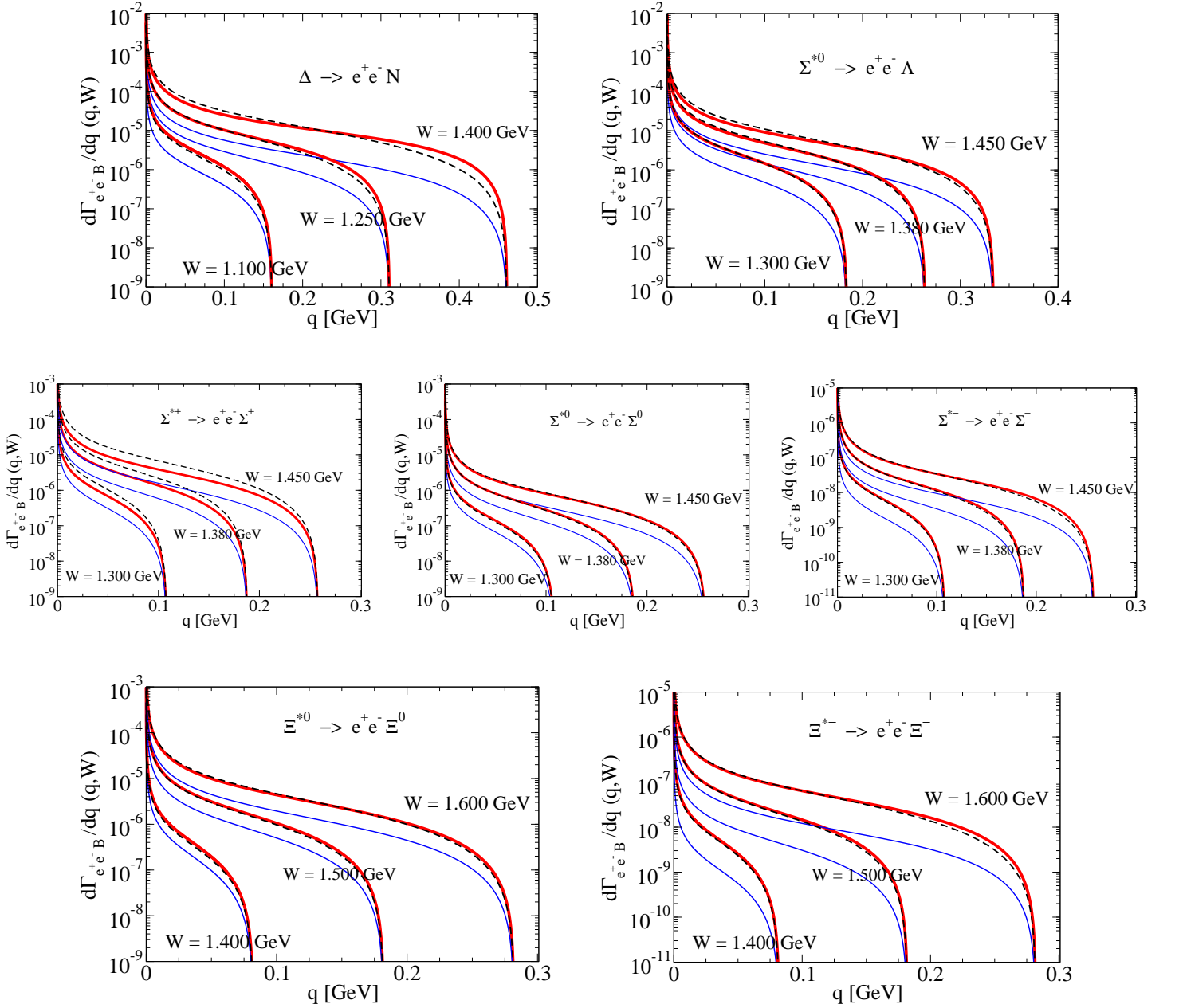

\centerline{\vspace{0.3cm} }
\centerline{
\mbox{
\includegraphics[width=3.1in]{DGamma-Nucleon-v3} \hspace{.6cm}
\includegraphics[width=3.1in]{DGamma-Lambda-v3} }}
\vspace{.7cm}
\centerline{
\mbox{
\includegraphics[width=2.4in]{DGamma-SigmaP-v3} \hspace{.15cm}
\includegraphics[width=2.4in]{DGamma-Sigma0-v3}  \hspace{.15cm}
\includegraphics[width=2.4in]{DGamma-SigmaM-v3} 
}}
\vspace{.7cm}
\centerline{
\mbox{
\includegraphics[width=3.1in]{DGamma-Xi0-v3} \hspace{.6cm}
\includegraphics[width=3.1in]{DGamma-XiM-v3} }}
\caption{\footnotesize{Dalitz decay rates 
$\frac{d \;}{d q}  \Gamma_{e^+e^-B}$
for different values of $W$.
Note the differences in the scales.
The thick solid lines represent our final result (bare plus meson cloud).
The thin solid lines represent the bare quark approximations.
The results of the constant form factor model 
($G_M(q^2,W) \to G_M(0,W)$) are indicated by the dashed lines.}}
\label{figDGamma}
\end{figure*}

\subsection{Radiative decays}

Our results for the radiative decay widths 
are a consequence of our estimate for $|G_M (0)|$ 
based on the dominance of magnetic dipole form factor.

The comparison between our estimates for  $|G_M (0)|$ 
and $\Gamma_{\gamma B} (M_{B'})$ are presented 
in Table~\ref{tableGM0}.
The experimental values for $|G_M (0)|$
are determined by Eq.~(\ref{eqGammaB}),
except for the $\Delta(1232) \to \gamma N$ decay,
where $G_M (0)$ is measured explicitly~\cite{PDG22}.
Notice that there are data only for 
the  $\Delta \to \gamma N$, 
$\Sigma^{*0} \to \gamma \Lambda$ 
and $\Sigma^{*+} \to \gamma \Sigma^+$ decays.
In the case of the $\Sigma^{\ast 0} \to \gamma \Sigma^0$, 
$\Sigma^{\ast -} \to \gamma \Sigma^-$ and 
$\Xi^{\ast -} \to \gamma \Xi^-$ decays, 
we present also the experimental limits 
(lower limit of the detectors).

In Table~\ref{tableGM0}, we display also our estimate without the inclusion
of the kaon cloud, $\left. G_M(0)\right|_{{\rm no} K} $,
and the contribution of the kaon cloud,  $\left. G_M(0)\right|_{K} $.
The enhancement of $\Gamma_{\gamma B}$ due to 
the kaon cloud can be inferred from the comparison of the two columns.
One can then conclude that the kaon cloud contributions
increase the radiative decays of the $\Sigma^{\ast}$ 
in about 25--30\%, except for the $\Sigma^{\ast 0}$,
and in about 50\% for the $\Xi^{\ast}$ decays.

Concerning the cases for which there are no measurements
of the Dalitz decay widths, we have a few remarks.

There is the expectation that, with the increasing 
of strangeness production in facilities such as 
HADES and PANDA~\cite{Salabura13,Lalik19,Rathod20a,Singh16},
that the $\Xi^{\ast 0} \to \gamma \Xi^0$ and 
$\Sigma^{\ast 0} \to \gamma \Sigma^0$ decay widths may be measured,
since their magnitudes are comparable to 
the $\Sigma^{\ast +} \to \gamma \Sigma^+$ decay width. 
The estimated $\Sigma^{\ast 0} \to \gamma \Sigma^{0}$ decay width 
is smaller than the previous three cases, but  we recall that $SU(3)$ 
and $U$-spin symmetry based calculations (see Sec.~\ref{secUspin}),
suggest that its magnitude is about one fourth 
of the $\Sigma^{\ast +} \to \gamma \Sigma^{+}$ decay width,
a reduction of less than an order of magnitude.

As for the  $\Sigma^{\ast -} \to \gamma \Sigma^{-}$ decay, 
for which we may expect a very small value
[zero in the exact $SU(3)$ symmetry limit].
One notices, however, that our model estimate, $|G_M (0)| \simeq 0.3$,
is very close to the experimental limit ($< 0.8$).
Finally, we point out that the estimated $\Xi^{\ast -} \to \gamma \Xi^-$ 
decay width has a magnitude close to the $\Sigma^{\ast -} \to \gamma \Sigma^{-}$ 
decay width.
A small improvement of the experimental precision 
may be sufficient to measure the  $\Sigma^{\ast -}$ and  $\Xi^{\ast -}$ 
radiative decays.


\subsection{Dalitz decay rates}

The derivatives of the Dalitz decay 
($B' \to e^+ e^- B$) width   
(Dalitz decay rates) for the possible decays 
of decuplet baryons to octet baryons 
are presented in Fig.~\ref{figDGamma}, for several values of $W$, 
near the decuplet baryon pole mass.
 
Notice that, although we use the sub-index $e^+ e^- B$,
these functions can be used for the 
$B' \to \mu^+ \mu^- B$ decay rates 
if we replace the range $2 m_e \le q  \le (W -M_B)$ 
by $2 \mu \le q  \le (W -M_B)$, where 
$2m_e \simeq 10^{-3}$ GeV and  $2 \mu \simeq 0.2$ GeV.

In Fig.~\ref{figDGamma}, we represent the full result
(bare plus meson cloud) by the thick lines
and the result of the bare contribution by the thin lines.
It is clear from the graphs that,
if we exclude the meson cloud contributions
(thin lines, bare contributions) we 
underestimate the final results by about 1 order of magnitude.

For the discussion about the impact
of including the $q^2$ dependence of the form factors 
on the Dalitz decay processes, 
we include also the estimates of the Dalitz decay rates
based on the QED approximation
(which neglects the $q^2$ dependence on $G_M$), 
when we replace $|G_M (q^2,W)|$ by $|G_M (0,M_{B'})|$
(value of $G_M$ at the pole).
We refer to this result also as the constant form factor model.
This result is represented by the dashed line.
In the cases where $|G_M (0,M_{B'})|$ are known,
$\Delta \to e^+ e^- N$,
$\Sigma^{*0} \to  e^+ e^- \Lambda$,  
and $\Sigma^{*+} \to  e^+ e^- \Sigma^+$, 
we use the experimental values of $|G_M (0,M_{B'})|$.
In the remaining cases the result of
the constant form factor model is determined 
by our value for $|G_M (0,M_{B'})|$.

Notice that, the full estimate (thick lines) and 
the constant form factor model (dashed lines)
are very close for the 
$\Sigma^{\ast 0}$, $\Sigma^{\ast -}$,  $\Xi^{\ast 0}$ 
and  $\Xi^{\ast -}$ decays, particularly 
for lower values of $q^2$, for all values of $W$.
From this this property, we can anticipate
that the results for the Dalitz di-electron decay widths
from our model (with $q^2$ dependence)
and from the constant form factor model may be very close.

In the case of the $\Sigma^{*+} \to  e^+ e^- \Sigma^+$,  
the difference between our estimate and 
the constant form factor model is in part a the consequence of the different estimates
for $|G_M (0,M_{\Sigma^{\ast}})|$.
Our calculation is below the experimental value in about
1.4 standard deviations (see Table~\ref{tableGM0}).
A similar observation is valid, but
with a smaller impact for the $\Sigma^{*0} \to  e^+ e^- \Lambda$ decay.

In the case of the $\Delta \to e^+ e^- N$,
and $\Sigma^{*0} \to  e^+ e^- \Lambda$ decays,
one can notice a deviation from the constant
form factor model when $W$ increases.
This effect is a consequence of 
the dependence of $|G_M|$ on $W$,
since the  $W$ dependence is absent 
in the constant form factor model.

\begin{figure*}[t]
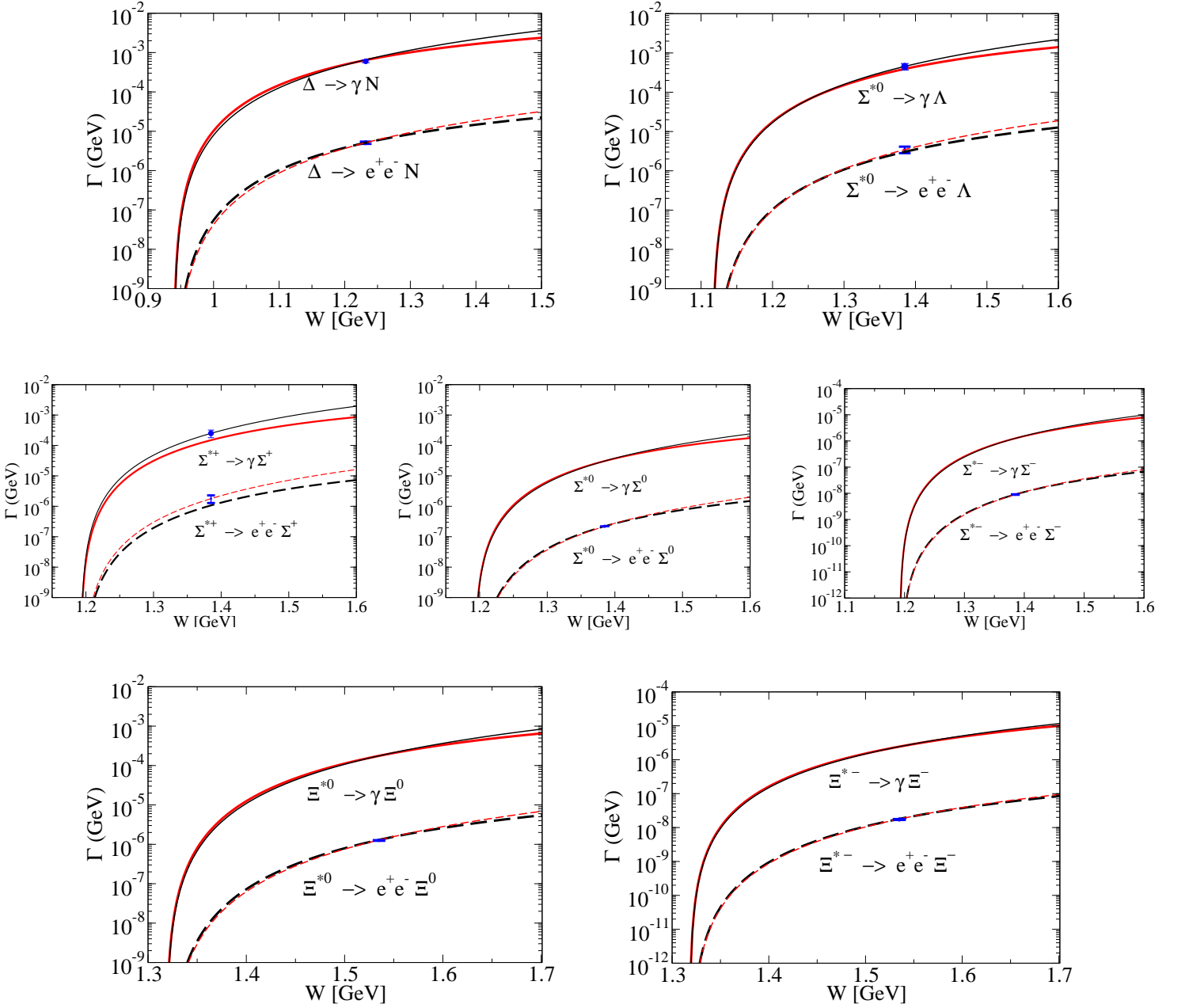

\centerline{\vspace{0.5cm}  }
\centerline{
\mbox{
\includegraphics[width=3.1in]{Gamma-Nucleon-v4} \hspace{.6cm}
\includegraphics[width=3.1in]{Gamma-Lambda-v3} }}
\vspace{.7cm}
\centerline{
\mbox{
\includegraphics[width=2.4in]{Gamma-SigmaP-v3}
\hspace{.3cm}
\includegraphics[width=2.4in]{Gamma-Sigma0-v3}
\hspace{.3cm}
\includegraphics[width=2.4in]{Gamma-SigmaM-v3} 
}}
\vspace{.7cm}
\centerline{
\mbox{
\includegraphics[width=3.1in]{Gamma-Xi0-v3} \hspace{.6cm}
\includegraphics[width=3.1in]{Gamma-XiM-v3} }}
\caption{\footnotesize{
Electromagnetic and Dalitz decay widths for all 
the decuplet decays in terms of $W$.
The thick lines represent our model.
The thin lines represent the constant form factor model.
The blue bullets represent the $\Gamma_{B^\prime \to \gamma B}$ data 
from Table~\ref{tableGM0}.
The pair of horizontal lines 
represent the limits of the $\Gamma_{B^\prime \to e^+ e^- B}$
estimate based on the constant form model,
presented in Table~\ref{tableDalitz}.}}
\label{figGamma}
\end{figure*}

\subsection{Dalitz decay widths}

The results of the integration of the Dalitz decay rates 
in $q$, in terms of $W$, the Dalitz decay widths, 
are presented in Fig.~\ref{figGamma}.
We use dashed lines to represent $\Gamma_{e^+e^-B} (W)$.
The thick lines represent our model and
the thin lines represent the constant form factor model.
The bands in the graphs for $\Gamma_{e^+e^-B} (W)$
indicate the result at the pole ($W=M_{B'}$).
In the cases the value of $G_M(0,M_{B'})$ are known, 
we include the window of variation associated with the estimate
associated with the constant form factor model.

For the first three decays 
($\Delta \to e^+ e^- N$, $\Sigma^{*0} \to e^+ e^- \Lambda$
and  $\Sigma^{*+} \to e^+ e^- \Sigma^+ $), 
we can notice that the constant form factor model predicts larger
contributions than our model.
This result is mainly a consequence of the reduction
of $|G_M (q^2, W) |$ when $W$ increases, in our model.
In the case of  $\Sigma^{*+} \to e^+ e^- \Sigma^+ $, the larger 
difference between estimates is a consequence of 
our result for $|G_M(0,W)|$, discussed already.
As for the remaining decays, we can notice that 
the two estimates are very close, even when $W$ increases.

\begin{table*}[t]
\begin{tabular}{l  c c c  c }
\hline
\hline
 Decay   &  BR$(\gamma B)$ &  $\tilde \Gamma_{e^+ e^- B}$ (keV)  &   $\Gamma_{e^+ e^- B}$ (keV) & BR$(e^+ e^- B)$ \\
\hline
\hline
$\Delta \to e^+ e^- N$              & 0.54$\times10^{-2}$   &  $5.11\pm0.37$ &  5.10 & 4.4$\times 10^{-5}$\\[.25cm]     
$\Sigma^{*0} \to  e^+ e^- \Lambda$   &  1.25$\times10^{-2}$   &  $3.47\pm0.65$ &   3.04  & 8.4$\times 10^{-5}$ \\[.25cm]    
$\Sigma^{*+} \to  e^+ e^- \Sigma^+$   &  0.69$\times10^{-2}$ &  $1.78\pm0.50$ &  1.07 & 3.0$\times 10^{-5}$ \\           
$\Sigma^{*0} \to  e^+ e^- \Sigma^0$   &   {\bf 0.086}$\times 10^{-2}$ &  {\bf 0.220} &  0.220 & 0.61$\times 10^{-5}$\\  
$\Sigma^{*-} \to  e^+ e^- \Sigma^-$   &   {\bf 0.0036}$\times 10^{-2}$ 
  & {\bf 0.0090} &  0.0091 &  0.023$\times 10^{-5}$\\[.25cm]  
$\Xi^{*0} \to  e^+ e^- \Xi^0$   &  {\bf 1.89}$\times 10^{-2}$   & {\bf 1.253} & 1.265 & 14$\times 10^{-5}$ \\    
$\Xi^{*-} \to e^+ e^- \Xi^-$   &   {\bf 0.023}$\times 10^{-2}$  & {\bf 0.0173}  &  0.0175  & 0.18$\times 10^{-5}$\\[.1cm]   
\hline
\hline
\end{tabular}
\caption{\footnotesize
Decuplet baryon Dalitz di-electron decay widths using 
the constant form factor model ($\tilde \Gamma_{e^+ e^- B}$)
and using $G_M(q^2,M_{B'})$   ($\Gamma_{e^+ e^- B}$).
The values in boldface are the estimates based
on our values for $G_M(0)$ from Table~\ref{tableGM0}. }
\label{tableDalitz}
\end{table*}

The general conclusion is that the Dalitz decay widths 
are sensitive to the values of $W$ (see Fig.~\ref{figGamma}).
This observation is important, because,
one can obtain larger values than the results 
at the pole ($\Gamma_{e^+ e^- B}(M_{B'})$) 
when the measurements are performed for larger values of $W$ ($W > M_{B'}$)
or smaller values when $W$ is smaller than $M_{B'}$.
This effect can be exemplified looking into the HADES measurements
of the $\Delta (1232)$ Dalitz decay.
Those measurements were performed for $W \simeq 1.39$ GeV,
a bit above the physical mass~\cite{HADES17}.
Notice that the increasing of $W$ implies also 
an increasing of the limit of the dilepton momentum
$q= W-M_N$ included in the integration.

In Fig.~\ref{figGamma},  we present also our results
for the radiative decay  widths in terms of $W$ (thick solid lines).
The data points close to the solid lines represent
the known data for $\Gamma_{\gamma B} (W)$ at the pole ($W=M_{B'}$).
The estimates from the constant form factor model
are represented by the thin solid lines.
In general, one can notice that the constant form factor model
has larger values than our model estimate
above the decuplet baryon mass pole.
This happens because the value of $|G_M(0,W)|$
is reduced when $W$ increases (see Fig.~\ref{figGM2}).
The exception is the $\Sigma^{*+} \to e^+ e^- \Sigma^+ $ case,
where we underestimate the QED value,
because we also underestimate $G_M$.

\subsection{Dalitz di-electron decay widths at the pole}

Our estimates for the Dalitz decay widths at the pole ($W= M_{B'}$) 
are presented in Table~\ref{tableDalitz}.
For the comparison with the constant form factor model, 
we include also the result of the calculation 
$|G_M(q^2, M_{B'})| \to |G_M(0, M_{B'})| \equiv |G_M(0)|$,
labeled as $\tilde \Gamma_{e^+ e^- B}$.
To determine $|G_M(0, M_{B'})|$, we use the 
procedure described in the previous sections.
The estimates based on our model for $|G_M(0, M_{B'})|$
are presented in Table~\ref{tableGM0} in boldface.
For discussion, we include also the 
$B^\prime \to \gamma B$ branching ratios, estimated by the data,
or by our model (last four cases).
In the cases in which $|G_M(0)|$ is determined by the data, 
we include also the estimated error based on the quadratic relation 
$\Gamma_{e^+ e^- B} \propto |G_M(0)|^2$
or $\Gamma_{e^+ e^- B} \propto \Gamma_{\gamma B}$.
The errors are large (larger than 5\%),
because the experimental determinations of 
$\Gamma_{\gamma B}$ are affected by significant errors,
even when $|G_M(0)|$ is well determined, 
as a consequence
of the quadratic relation mentioned above.\footnote{The relative error 
in $\Gamma_{\gamma B}$ is twice the relative error 
in  $|G_M(0)|$. 
Thus, if $|G_M(0)|$ is determined with a precision of 10\%,
$\Gamma_{\gamma B}$ is known with a precision of 20\%.}
The estimates of the constant form factor model
must then be taken with care due to the errors associated with $|G_M(0)|$.
The estimate of $\tilde \Gamma_{e^+ e^- B}$ is very sensitive to the 
value used for $|G_M(0)|$ or equivalently by $\Gamma_{\gamma B}$.

We can now discuss our estimates for the Dalitz di-electron decay widths.
From the results for the last four decays 
($\Sigma^{\ast 0,-}$ and $\Xi^{\ast 0,-}$), 
we conclude that there are no significant 
differences to the estimates based on a constant form factor.
The differences between estimates are 1\% at most.
Since $|G_M(0,M_{B'})|$ was also estimated by our model,
the conclusion is that the $q^2$ dependence 
on the form factors is not relevant 
in the integration of the Dalitz di-electron decay rates. 
These results are expected from the results
shown in Fig.~\ref{figDGamma} for the Dalitz decay rates.

Our result for the $\Delta(1232)$
compares well with the unique measurement at HADES of
the $\Delta(1232)$ Dalitz decay at the moment.
The HADES result is $4.90 \pm 0.83$ keV~\cite{HADES17}.
The new calculation (5.10 keV), with the extra kaon cloud (4\% effect)
is still compatible with HADES.

When compared with our previous work~\cite{DecupletDalitz},   
where the kaon cloud contributions were not taken into account, 
we conclude that the kaon cloud increases 
the Dalitz decay width in 27\% for 
$\Sigma^{\ast 0 } \to e^+ e^- \Lambda$ decay, 
30\%--40\% for the $\Sigma^{\ast} \to e^+ e^- \Sigma$ decays,
and 50\%--60\% for the $\Xi^{\ast} \to e^+ e^- \Xi$ decays.
The more significant enhancements in absolute values happen
for $\Sigma^{\ast 0 } \to e^+ e^- \Lambda$,
 $\Sigma^{\ast +} \to e^+ e^- \Sigma^+$ 
and  $\Xi^{\ast 0} \to e^+ e^- \Xi^0$ decays.
The result of  $\Xi^{\ast 0} \to e^+ e^- \Xi^0$ decay
is relevant, because of its magnitude and by the expectation
that the $\Xi^\ast$ Dalitz decays 
may be measured for the first time, 
as a consequence of the increasing
of the  production of hyperons with two
strange quarks at HADES~\cite{Lalik19,Rathod20a,Singh16,HADES21a,Xu22a}.

Concerning the $\Delta \to e^+ e^- N$,
$\Sigma^{\ast 0}  \to  e^+ e^- \Lambda$ 
and $\Sigma^{\ast +}  \to e^+ e^- \Sigma^+$ decays, 
our estimates are close to the estimates of the constant 
form factor model for the first two cases,
and underestimate the result for $\Sigma^{\ast +} \to e^+ e^- \Sigma^+$ decay.
The underestimation for the last case is a consequence of
our underestimation of $G_M$ near $q^2=0$, as explained in 
the previous sections.
We recall that the constant form factor is determined 
in these cases by the experimental value for the radiative decay.
The estimated errors associated with the constant form factor model
are then related to the present precision for
the radiative decays $\Gamma_{\gamma B}$.

The improvement in accuracy in the measurements of the radiative decays
can help to discriminate the differences between our model 
and the QED estimate.
If  this is not possible in a near future, one should
look for the results for the Dalitz decay rates, 
$\frac{d \;}{d q} \Gamma_{e^+ e^- B} (q,W)$, as in Fig.~\ref{figDGamma},
to detect differences between the two calculations.

The results for Dalitz di-electron decay rates depend on the 
explicit value of $W$, 
as one can conclude from the analysis of Fig.~\ref{figGamma}.
It is worth noticing that given a model for $|G_T (q^2,W)|$, the result 
is closer to the constant form factor model estimate when $W-M_B$ is small.
This happens because in the calculations 
$|G_T (q^2,W)|^2$ is multiplied by the factor
$\left( 1 - q^2/(W - M_B)^2\right)^{3/2}$,
according to Eq.~(\ref{eqGammaG}),
leading to a suppression of the integrand function near $q=W-M_B$.
Thus,  the integral of the Dalitz decay rate 
(\ref{eqGammaInt}) is dominated by  $|G_T (q^2,W)|$ near $q^2=0$
when $W-M_B$ is small, and the estimate based 
on the constant form factor is a good approximation.

The results for the $\Sigma^{\ast 0} \to e^+ e^- \Lambda$ 
and  $\Sigma^{\ast +} \to e^+ e^- \Sigma^+$ decays are very important 
because they show that the kaon cloud contributions enhance  
the form factors in the timelike region,
increasing the estimates which take only the pion cloud into account.
This observation is particularly relevant for the 
diagram (b) in the  $\Sigma^{\ast 0} \to e^+ e^- \Lambda$ decay.

In a recent work, a HADES feasibility study 
based on the constant form factor, 
 estimated the  $\Sigma^{*0} \to e^+ e^- \Lambda$
Dalitz decay width as $5.04 \pm 0. 70$ keV~\cite{HADES21a}.
This result is a consequence of the assumption of the 
constant form factor with $|G_M (0)| \simeq 3.49$
(estimate based on BR= $1.4 \times 10^{-4}$
and $\Gamma_{\gamma B} = 504$ keV) 
and an approximation in the numerical calculation 
(overestimate of 10\%).
An estimate with $|G_M (0)| \simeq 3.49$ leads
to a Dalitz decay of 3.90, about 30\%
below HADES feasibility estimate.
The difference can also be due to different regions 
of $W$ (average of the decuplet baryon mass) in the calculations.

The $\Sigma^{*0} \to e^+ e^- \Lambda$ decay widths
can also be compared with estimates from 
chiral perturbation theory combined 
with dispersion relations (3.0--3.4 keV)~\cite{Junker20a}.

More recently, the decuplet baryon Dalitz decays 
were estimated with an $SU(3)$ symmetry breaking model~\cite{Xu22a}.
The authors used a linear approximation to
the transition form factors 
$G (q^2) \simeq G(0) \left( 1 + \sfrac{1}{6} q^2 \left< r \right>^2  \right)$ and 
calculated the interval values for the 
the Dalitz decay widths associated with
the range $0 \le \left< r \right>^2 \le 1 $ fm$^2$.
Their results are close to our results when 
we include the interval of variation 
(between 20\% and 40\%), except for 
the $\Sigma^{*0} \to e^+ e^- \Lambda$ decay width, 
$2.25\pm0.53$ keV, a bit below our new estimate (3.04 keV).

From the analysis of the branching ratios it is interesting to notice 
that the largest value for the  Dalitz di-electron decay is for 
the $\Xi^{\ast 0} \to e^+ e^- \Xi^0$ decay, 
with a magnitude a bit above the $\Sigma^{\ast 0} \to e^+ e^- \Lambda$ decay.
This estimate provides an additional motivation
for the experimental determination of the radiative 
and Dalitz di-electron decays.
Another point of interest is the similarities of the magnitudes 
of the branching ratios for $\Xi^{\ast -} \to e^+ e^- \Xi^-$
and $\Delta \to e^+ e^- N$ decays, 
as well as $\Sigma^{\ast 0} \to e^+ e^- \Lambda$ and $\Sigma^{\ast +} \to e^+ e^- \Sigma^+$ decays.
Notice that three of these decays
are associated with channels where the radiative decay has been measured,
while the $\Xi^{\ast -}$ decay is strongly suppressed 
in the $SU(3)$ symmetry limit
[the non-zero value is a consequence of $SU(3)$ symmetry breaking].


\subsection{Dalitz di-muon decay widths at the pole}

Using the formalism of the Dalitz dilepton decays,  
we can also calculate the Dalitz decay widths into 
muons ($B' \to \mu^+ \mu^- B$), 
based on Eqs.~(\ref{eqGammaP2})--(\ref{eqGammaInt2}).
The results for the Dalitz di-muon decay widths and the branching ratios 
estimated using our values and the Particle Data Group (PDG)  experimental values
for the total decay widths are given in Table~\ref{tableDalitz-mu}.

In the case of the di-muon decays, 
the $\Sigma^* \to \Sigma$ decays are forbidden because 
$2 \mu > M_{B'}-M_B$
($\mu$ is the mass of the muon).
The forbidden character of the  $\Sigma^*$ decays 
can be inferred from the results shown in
Fig.~\ref{figDGamma} for $W \simeq 1.38$ GeV.
Notice that the upper limit $q \simeq W -M_\Sigma \simeq 0.19$ GeV
is below $2 \mu \simeq 0.21$ GeV.

The main points of interest are the magnitudes of the decay widths 
and branching ratios. 
Numerically, the $B' \to \mu^+ \mu^- B$ decay widths
are better expressed in units of eV.
The more relevant cases are the $\Delta$ and $\Sigma^{\ast 0} \to \Lambda$ 
decays with branching ratios of about 2 orders of magnitude
smaller than the di-electron case ($10^{-7}$ versus $10^{-5}$).
Branching ratios of $10^{-7}$ are within the accuracy of LHCb experiments~\cite{LHCb} 
and are expected to be observed in BESIII and LHCb in a near future~\cite{Xu22a}.

\begin{table}[t]
\begin{tabular}{l  c c}
\hline
\hline
 Decay   &  $\Gamma_{\mu^+ \mu^- B}$ (eV)  &   BR$(\mu^+ \mu^- B)$  \\
\hline
\hline
$\Delta \to \mu^+ \mu^- N$ & 36.3 &  0.31$\times 10^{-6}$ \\[.25cm]
$\Sigma^{*0} \to  \mu^+ \mu^- \Lambda$  &  9.70 &  0.27$\times 10^{-6}$\\[.25cm]
$\Xi^{*0} \to  \mu^+ \mu^- \Xi^0$   & 1.9$\times 10^{-3}$ &   0.21$\times 10^{-9}$ \\
$\Xi^{*-} \to \mu^+ \mu^- \Xi^-$   &  2.8$\times 10^{-5}$ &  2.8$\times 10^{-12}$\\
\hline
\hline
\end{tabular}
\caption{\footnotesize
Decuplet baryon Dalitz di-muon decay widths  
and branching ratios.}
\label{tableDalitz-mu}
\end{table}

The Dalitz decay widths into di-muons 
for $\Delta \to N$ and $\Sigma^{\ast 0} \to \Lambda$ 
are about 2 orders of magnitude smaller than
the Dalitz decay widths into di-electrons.
As for the $\Xi^{\ast } \to \mu^+ \mu^- \Xi$ decay widths, they  
are about $10^{-6}$ smaller (6 orders of magnitude)
than the decay widths into electrons.
This result is mainly a consequence of the 
phase space~\cite{Xu22a}, which is strongly suppressed 
for muons~\cite{Xu22a}.
The range of integration of $q$ in the cases of muons 
is $[0.211,0.215]$ GeV.
Notice in this respect that, the calculation of the $\Xi^\ast$
di-muon decays requires a value of $W= 1.533$ GeV,
between the values of $W$ displayed in Fig.~\ref{figDGamma}.

Our estimates are compatible with the estimates
from Xu {\it et al.}~\cite{Xu22a} for the $\Delta \to  \mu^+ \mu^- N$
and $\Sigma^{*0} \to  \mu^+ \mu^- \Lambda$ decays.
Also for the cascades, we obtain estimates consistent  
with Ref.~\cite{Xu22a}, within their large error bars.

\begin{table}[t]
\begin{tabular}{l  c c c }
\hline
\hline
 Decay   &  BR($\gamma B$) &  BR($e^+ e^- B$)  &   BR($\mu^+ \mu^- B$)  \\
\hline
\hline
$\Delta \to \ell^+ \ell^- N$          & 0.54$\times10^{-2}$   &  $4.4 \times 10^{-5}$ &  $3.1 \times 10^{-7}$ \\[.25cm]     
$\Sigma^{*0} \to  \ell^+ \ell^-  \Lambda$   &  1.25$\times10^{-2}$   &  $8.4 \times 10^{-5}$  & $2.7 \times 10^{-7}$ \\[.25cm]    
$\Sigma^{*+} \to  \ell^+ \ell^- \Sigma^+$   &   0.69$\times10^{-2}$ &   $3.0 \times 10^{-5}$  &   \\           
$\Sigma^{*0} \to  \ell^+ \ell^- \Sigma^0$   &   {\bf 0.086}$\times 10^{-2}$ &  $6.1 \times 10^{-6}$  &  \\  
$\Sigma^{*-} \to  \ell^+ \ell^- \Sigma^-$   &   {\bf 0.0036}$\times 10^{-2}$
  & $2.3 \times 10^{-7}$  &   \\[.25cm]  
$\Xi^{*0} \to  \ell^+ \ell^- \Xi^0$   &  {\bf 1.89}$\times 10^{-2}$   & $1.4\times 10^{-4}$ &   
$2.1\times 10^{-10}$ \\    
$\Xi^{*-} \to \ell^+ \ell^- \Xi^-$   &   {\bf 0.023}$\times 10^{-2}$  & $1.8\times 10^{-6}$  &   $2.8\times 10^{-12}$ \\[.1cm]  
\hline
\hline
\end{tabular}
\caption{\footnotesize  Summary of branching ratios. 
The results in boldface for BR($\gamma B$) are based on the estimates
of the present model.}
\label{tableDalitz3}
\end{table}

\subsection{Summary of Dalitz dilepton decays}

In Table \ref{tableDalitz3}, we summarize our knowledge of 
radiative, di-electron and di-muon decays, in terms of the branching ratios.
The last four radiative decays are estimated based
on our own calculations for the magnetic form factor 
(indicated in boldface).

The results from Table~\ref{tableDalitz3}
allow us to compare the magnitudes of the branching ratios
and discuss the expectations of measurements in
future experiments. 
Our calculations indicate that the $\Xi^{*0} \to  e^+ e^- \Xi^0$
decay has the largest branching ratio ($\sim 10^{-4}$)
followed by the $\Delta$, $\Sigma^{*0} \to  e^+ e^- \Lambda$ 
and $\Sigma^{*+}$ decays (branching ratios $\sim 10^{-5}$).
The $\Sigma^{*0} \to  e^+ e^- \Sigma^0$  and $\Xi^{*-}$ decays 
have branching ratios of the order $10^{-6}$.
The last result shows that the $\Xi^{*-}$ Dalitz decay,
suppressed in the exact $SU(3)$ symmetry, has a branching ratio
just an order of magnitude below the  
$\Sigma^{*+} \to  e^+ e^- \Sigma^+$ decay.
The last decay is expected to be observed in a near future~\cite{HADES21a}.
Only the $\Sigma^{*-}$ decay has reduced branching ratio ($\sim 10^{-7}$).

As for the di-muon Dalitz decays,
the $\Delta$ and $\Sigma^{*0} \to  \mu^+ \mu^- \Lambda$ decays
have branching ratios of the order of $10^{-7}$, 
only 2 orders of magnitude below the measured 
$\Delta(1232)$ Dalitz decay width at HADES,
and the HADES/PANDA feasibility study for 
$\Sigma^{*0} \to e^+ e^- \Lambda$~\cite{HADES17,HADES21a}.

With the possibility of creation of a large number of baryons 
with strange quarks at HADES, PANDA, BESIII, and LHCb,
one can expect to measure in a near future 
di-electron Dalitz decay widths 
for the decays $\Sigma^{*0} \to \Lambda$, 
$\Sigma^{*+} \to \Sigma^+$,   $\Sigma^{*0} \to \Sigma^0$,
$\Xi^{*0} \to \Xi^0$ and $\Xi^{*-} \to \Xi^-$,
as well as di-muon Dalitz decay widths 
for the $\Delta \to N$ and 
$\Sigma^{*0} \to \Lambda$ transitions.



\vspace{.25cm}

\section{Outlook and conclusions}
\label{secConclusions}

Most of the experimental studies about the electromagnetic structure 
of the baryons have been performed in the spacelike region.
Only recently, the access to the timelike region of the 
baryon transitions ($\gamma^\ast B \to B'$) became possible
in {\it direct} electromagnetic transitions 
with the measurement of the $\Delta(1232)$ di-electron 
Dalitz decay width at HADES.
Measurements of baryon radiative decays and 
Dalitz decays are very important for our understanding 
of the internal structure of nucleon resonances and hyperons.

The characteristics of the HADES and PANDA facilities 
indicate that measurements of Dalitz di-electron decays 
of nucleon resonances and hyperons can be measured at GSI 
in the near future.
There is also the expectation based on 
the estimated branching ratios that measurements 
of baryon Dalitz di-muon decays may be observed 
in facilities such as BESIII and LHCb.

From the possible electromagnetic decays of
decuplet baryons into octet baryons, 
only three radiative decays are known presently:
$\Delta(1232) \to \gamma N$, $\Sigma^0(1385) \to \gamma \Lambda(1116)$,
and  $\Sigma^+(1385) \to \gamma \Sigma^+(1193)$.
The present work suggests,  based on the
estimated magnitudes and improvement of the 
experimental conditions, that the 
$\Xi^0(1530) \to \gamma \Xi^0(1318)$ and $\Sigma^0(1385) \to \gamma \Sigma^0(1193)$ decays 
can be measured in the next few years.
There is also the possibility that the  
$\Sigma^-(1385) \to \gamma \Sigma^-(1193)$ decay,
forbidden in the case that the exact $SU(3)$ flavor symmetry holds,
may also be measured soon due to
the closeness of the estimated result with the present experimental limit.
We may also expect observations of the
$\Xi^-(1530) \to \gamma \Xi^-(1318)$ decay due 
to the similarities with the $\Sigma^-(1385) \to \gamma \Sigma^-(1193)$ decay.

From the feasibility studies of Dalitz di-electron decays 
at HADES and PANDA and the progress in the measurements 
of the decuplet baryon radiative decays,
one may expect that the 
$\Sigma^0(1385) \to e^+ e^- \Lambda(1116)$,
$\Sigma^+(1385) \to e^+ e^- \Sigma^+(1193)$
and $\Xi^0(1530) \to e^+ e^- \Xi^0(1318)$ decays
will be measured soon at HADES and PANDA.
The estimated magnitudes of the branching ratios
suggest also that the $\Delta(1232)$ and $\Sigma^0 (1385) \to \Lambda(1116)$
decays into di-muons may be measured at BESIII or LHCb.

In the present work, we study the role of the meson cloud
on the octet baryon to decuplet baryon transition form factors,
with particular focus on the kaon cloud, which is more relevant
to baryons with strange quarks.
This effect could in principle be studied in the spacelike region. 
In the spacelike region, however, our knowledge is restricted
to the $\gamma^\ast N \to \Delta(1232)$ transition,
which manifests a very weak dependence on the kaon cloud,
since the non-valence contributions
are dominated by the pion cloud processes.
In these conditions the impact of the kaon cloud 
can be better seen on the transition form 
factors related to the decuplet baryons 
$\Sigma(1385)$ and $\Xi(1530)$.
We conclude that the kaon cloud can enhance
the estimates of the $\Sigma(1385)$ radiative decays 
in about 27\% and the $\Xi(1530)$ radiative decays in about 50\%. 
We focus also our interest in the
$q^2$ dependence of the form factors
with impact on the Dalitz decay widths.
We conclude that the $q^2$ dependence 
of the form factors may be more relevant 
for the $\Sigma^0(1385) \to \Lambda(1116)$, and
$\Sigma^+(1385) \to \Sigma^+(1193)$ transitions.
As for the $\Sigma^{0,-}(1385) \to \Sigma^{0,-}(1193)$
and  $\Xi^{0,-}(1530) \to e^+ e^- \Xi^{0, -}(1318)$,
we conclude that the $q^2$ dependence 
is not relevant for the calculation of the 
Dalitz di-electron decay widths.

The final conclusion is that 
the present estimates may be used to guide the
experimental studies of baryon Dalitz decays,
and that our predictions may be tested soon 
at HADES, PANDA or other facilities.
The formalism used in the present work may be used
in future theoretical studies,
extending the  $SU(3)$ flavor sector 
(quarks $u$, $d$ and $s$)
with the inclusion of the heavy quarks $c$ or $b$.
Within this extension it will be possible to
calculate radiative and Dalitz decays of baryons with heavy quarks.

\begin{acknowledgments}
The authors acknowledge the support and warm hospitality of APCTP
(Asia Pacific Center for Theoretical Physics) during the Workshop (APCTP PROGRAMS 2023)
''Origin of Matter and Masses in the Universe: Hadrons in free space, dense nuclear medium,
and compact stars'', where important discussions and development were achieved
on the topic.
K.~T.~thanks the OMEG (Origin of Matter and Evolution of Galaxies) Institute at Soongsil
University for the supports in many aspects during the collaboration visit in Korea.
G.~R.~thanks to the LFTC group of the
Universidade Cruzeiro do Sul and Universidade Cidade de S\~ao Paulo for their 
hospitality during the visit in February of 2023.
G.~R.~was supported the Basic Science Research Program 
through the National Research Foundation of Korea (NRF)
funded by the Ministry of Education  (Grant No.~NRF–2021R1A6A1A03043957).
K.~T.~was supported by Conselho Nacional de Desenvolvimento
Cient\'{i}fico e Tecnol\'ogico (CNPq, Brazil), Processes No.~313063/2018-4 
and No.~426150/2018-0, and FAPESP Process No.~2019/00763-0, 
and his work was also part of the projects, 
Instituto Nacional de Ci\^{e}ncia e 
Tecnologia - Nuclear Physics and Applications 
(INCT-FNA), Brazil, Process No.~464898/2014-5, 
and FAPESP Tem\'{a}tico, Brazil, Process No.~2017/05660-0.
\end{acknowledgments}

\appendix

\section{Quark form factors}
\label{appQuarkFF}

\setcounter{figure}{0}
\renewcommand\thefigure{\thesection.\arabic{figure}}    

Motivated by the VMD mechanism, 
we use the following parametrization for the 
quark form factors 
$f_{i0}$ and $f_{i\pm}$ ($i=1,2$):
\ba
&&
\hspace{-1.2cm}
f_{1+}(q^2) = \lambda_q + (1 - \lambda_q) \frac{m_\omega^2}{m_\omega^2 - q^2} 
- c_+ \frac{M_h^2 q^2}{(M_h^2 - q^2)^2}, 
\label{eqF1p}\\
& &
\hspace{-1.2cm}
f_{1-}(q^2)  = \lambda_q + (1 - \lambda_q) \frac{m_\rho^2}{m_\rho^2 - q^2} 
- c_- \frac{M_h^2 q^2}{(M_h^2 - q^2)^2}, 
\label{eqF1m}\\
& &
\hspace{-1.2cm}
f_{10} (q^2) = \lambda_q + (1 - \lambda_q) \frac{m_\phi^2}{m_\phi^2 - q^2} 
- c_0 \frac{M_h^2 q^2}{(M_h^2 - q^2)^2}, 
\label{eqF10}\\
&&
\hspace{-1.2cm}
f_{2+}(q^2) =   \kappa_+ \left\{ 
d_+ 
 \frac{m_\omega^2}{m_\omega^2 - q^2} 
 + (1- d_+)  \frac{M_h^2}{M_h^2 - q^2}  \right\},
 \label{eqF2p} \\
&&
\hspace{-1.2cm}
f_{2-}(q^2) =   \kappa_- \left\{ 
d_- 
 \frac{m_\rho^2}{m_\rho^2 - q^2} 
+ (1- d_-)  \frac{M_h^2}{M_h^2 - q^2}  \right\}, 
\label{eqF2m}\\
&&
\hspace{-1.2cm}
f_{20}(q^2) =   \kappa_0 \left\{ 
d_0 
 \frac{m_\phi^2}{m_\phi^2 - q^2} 
+ (1- d_0)  \frac{M_h^2}{M_h^2 - q^2}  \right\}, 
\label{eqF20}
\ea 
where $m_\rho$, $m_\omega$, and $m_\phi$ 
represent the masses of the mesons 
$\rho$, $\omega$, and $\phi$, respectively.
The terms with $M_h$ correspond to an effective heavy vector meson
which parametrize the short range effects.
The value of $M_h$ is fixed as $M_h = 2 M_N$~\cite{Nucleon,Octet1}.
In the calculations, we use the 
approximation $m_\omega = m_\rho$ for simplicity.
We use the values $m_\rho= 0.771$ GeV
as discussed in Appendix~\ref{appPionCloud}
and $m_\phi= 1.020$ GeV.

In Eqs.~(\ref{eqF2p})--(\ref{eqF20}), 
$\kappa_\pm$ represent the isoscalar/iso-vector 
quark anomalous magnetic moments,
and $\kappa_0$ is the strange quark 
anomalous magnetic moment.
We use the parametrization derived from the studies
of the octet and decuplet baryons~\cite{Octet2,Omega}.
We take in particular $\kappa_- = 1.435$,  
$\kappa_+ = 1.803$, and $\kappa_0 = 1.462$.
To convert to the flavors $q=u,d,s$, 
one uses the following relations: 
 $\kappa_u = \frac{1}{4}(\kappa_+ + 3 \kappa_-)$,
$\kappa_d = \frac{1}{2}(2 \kappa_- - \kappa_+)$,  
and $\kappa_s= \kappa_0$~\cite{Nucleon,Omega}.


In Eqs.~(\ref{eqF1p})--(\ref{eqF10}), $\lambda_q$
is a parameter related to
the quark density number in deep inelastic scattering~\cite{Nucleon}.
The numerical value is $\lambda_q =1.21$.
The remaining parameters are 
$c_+= 4.160$, $c_-=1.160$, $c_0= 4.427$,
$d_+= d_- =-0.686$, and $d_0=-1.860$~\cite{Octet2,Omega}.

The expressions (\ref{eqF1p})--(\ref{eqF20}) are valid 
in the region \mbox{$q^2 < 0$}, when the
vector meson decay widths vanish, $\Gamma_v  \equiv 0$ ($v=\rho, \omega,\phi$).
For the extension of the quark form factors 
to the timelike region ($q^2 > 0$), we consider 
the replacement ($v=\rho, \omega, \phi$)
\ba
\frac{m_v^2}{m_v^2 -q^2} 
\to 
 \frac{m_v^2}{m_v^2 -q^2 - i m_v \Gamma_v (q^2)}. 
\ea
The decay width functions $\Gamma_v(q^2)$, which describe 
the dressing of the vector mesons in terms 
of the possible meson decay channels, are discussed next.

For the $\rho$ meson (isovector component) 
the decay width parametrizes
the width associated with the decay $\rho \to 2 \pi$
for a virtual $\rho$ with square four-momentum $q^2$.
This function can be expressed
as in Eq.~(\ref{eqGammaRho}),
due to the connection with the
pion electromagnetic form factor~\cite{Connell95}.

\begin{figure}[t]
\vspace{.5cm}  
\includegraphics[width=3.0in]{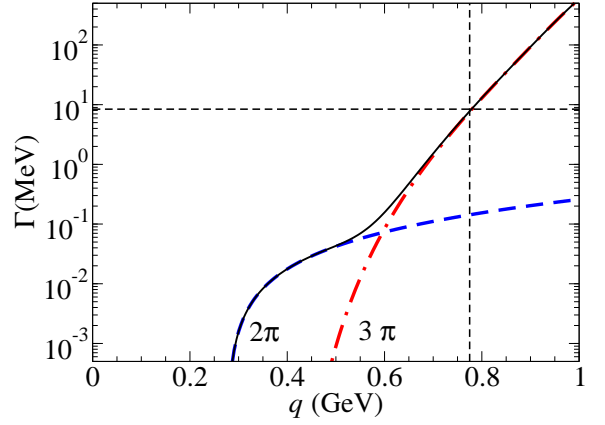}
\caption{\footnotesize
Function $\Gamma_\omega$ versus $q$.
The solid line represents the sum of the two channels. The 
short-dashed vertical and horizontal lines indicate, respectively, 
the $\omega$ mass point and
the $\omega$-physical width (8.4 MeV).}
\label{fig-Gamma-omega}
\end{figure}

As for the isoscalar channel, associated with the $\omega$ meson,
one needs to consider the combination 
of the decays $\omega \to 2\pi$ and $\omega \to 3\pi$.
Following our work on the $N(1520)$ Dalitz decay~\cite{N1520TL},
we decompose  
\ba
\Gamma_\omega (q^2) = 
\Gamma_{2 \pi} (q^2) + \Gamma_{3 \pi} (q^2),
\ea
where the first term parametrize the decay  $\omega \to 2\pi$ 
and the second term  parametrize the decay  $\omega \to 3\pi$.
The expression for $\Gamma_{2 \pi} (q^2) $ is similar 
to $\Gamma_{\rho} (q^2) $ except for the strength~\cite{N1520TL,Muhlich}.
For the decay  $\omega \to 3\pi$, 
we consider a model based on the  
process $\omega \to \rho \pi \to 3 \pi$,
where the intermediate $\rho$ decays into two pions~\cite{Muhlich}.
We do not write down here the expressions for $\Gamma_{2 \pi}$ 
and $\Gamma_{3 \pi}$, since they can be found in Ref.~\cite{N1520TL},
except for the expression for $ \Gamma_{2 \pi} (q^2)$ 
where we use the factor $\frac{m_\omega}{\sqrt{q^2}}$ 
instead of $\frac{m_\omega^2}{q^2}$
and replace the $\pi \rho$ coupling from 
$g^\prime = 10.6$ MeV to $g^\prime = 13.86$ MeV,
in order to reproduce $\Gamma_{3\pi} (m_\omega^2) = 7.6$ MeV~\cite{N1520TL}.
The numerical results for the function $\Gamma_\omega (q^2)$
are presented in Fig.~\ref{fig-Gamma-omega}.

Finally, for the $\phi$ decay width, we consider 
the simplified  parametrization,
\ba
\Gamma_\phi(q^2) = \Gamma_\phi^0 
\frac{m_\phi}{\sqrt{q^2}} 
\left(\frac{q^2 - 4 m_K^2}{m_\phi^2- 4 m_K^2} \right)^{3/2}
\theta(q^2 - 4 m_K^2), \nonumber\\
\label{eqGammaPhi}
\ea
where $\Gamma_\phi^0 = 4.23 \times 10^{-3}$ GeV, 
and $m_K$ is the kaon mass ($m_K \simeq 0.5$ GeV).
Equation (\ref{eqGammaPhi}) describes the $\phi \to 2 K$ 
($K$ is the kaon) under the assumption that it is the dominant 
decay of the $\phi$.
According to PDG $\phi$ decays predominantly
on 2$K$ (49\%) and $K_S^0 K_L^0$ (34\%)~\cite{PDG22}.

For the range of the calculation  of the present 
work ($W < 2$ GeV)
the regularization of the $\phi$ pole is not 
very relevant, since $m_\phi^2 \simeq 1$ GeV$^2$ $\gg q^2$.
The singularities associated with the 
 $\phi$ meson appear, then only for $W \ge M_B + m_\phi > 2.1$ GeV.


\section{Parametrization of the pion cloud contribution}
\label{appPionCloud}

\setcounter{figure}{0}
\renewcommand\thefigure{\thesection.\arabic{figure}}

\begin{figure*}[t]
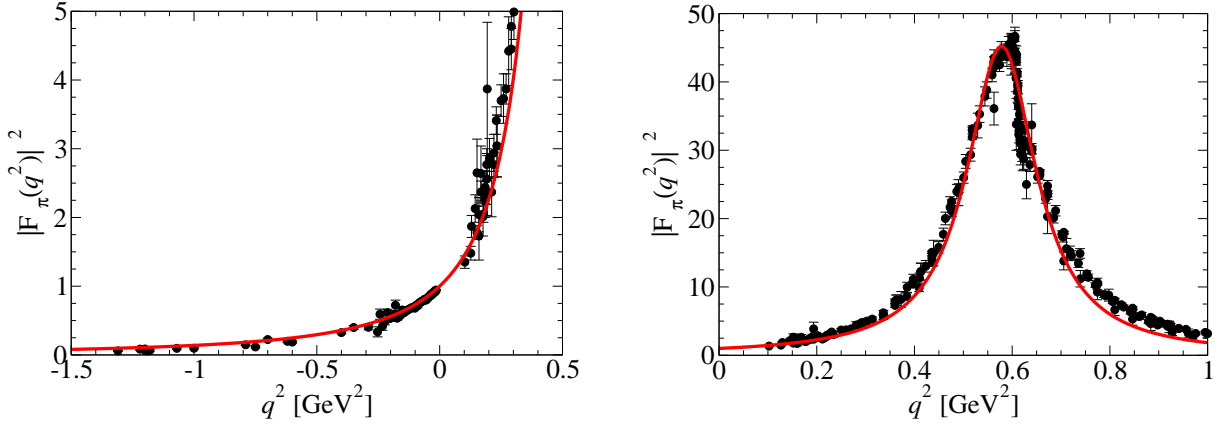

\vspace{.5cm}
\centerline{
\mbox{\includegraphics[width=3.0in]{Fpi2-v2} \hspace{.6cm}
\includegraphics[width=3.0in]{Fpi2-v1b} }}
\caption{\footnotesize 
Comparison of the best fit (\ref{eqFpi}) with the spacelike (left panel) 
and timelike (right panel) $|F_\pi (q^2)|^2$ 
data~\cite{PionFF_data_TL,PionFF_data_SL}.
The parameters associated with the best fit 
are $m_\rho = 0.771$ GeV and $\Gamma_\rho^0 = 0.117$ GeV.}
\label{fig-FitFpi2}
\end{figure*}

\begin{figure}[t]
\vspace{.2cm}
\includegraphics[width=3.0in]{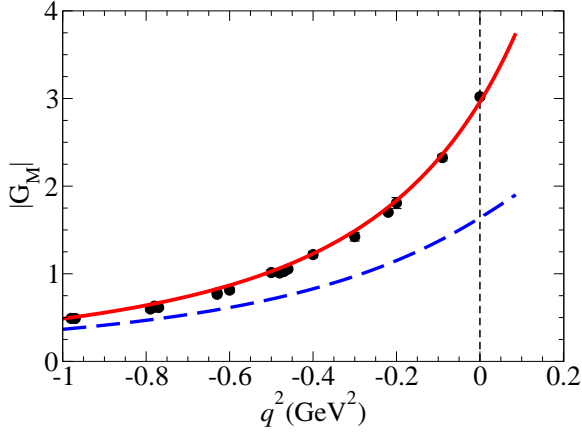}
\caption{\footnotesize
Results from the covariant spectator quark model for
$|G_M (q^2,M_\Delta)|$ relative to the $\gamma^\ast N \to \Delta$ transition (solid line). 
The valence and pion cloud contributions 
are combined according to Eq.~(\ref{eqGM-total}).
The data are from Refs.~\cite{PDG10,GM-Data}.
The dashed line is the contribution from the core (valence quarks).}
\label{fig-GM-total}
\end{figure}

We discuss here the pion cloud parametrization (\ref{eqGMpi}) 
and the function (\ref{eqFpi}).
We fit the function $|F_\pi (q^2)|^2$ to the data of respective regions of
spacelike~\cite{PionFF_data_SL} and timelike~\cite{PionFF_data_TL}.

Since we focus on the isovector component
of  $F_\pi (q^2)$ ($\rho \to \pi \pi$), 
we do not expect an exact description near the peak due
to the overlap of the data with the 
isoscalar component ($\omega \to \pi \pi$).
In order to optimize a description of 
lower $q^2$ region (below the peak) and 
also to  avoid the contamination by the excited states of the $\rho$, 
we restrict the fit to the $q^2 < 0.8$ GeV$^2$ data.
The parameters associated with the best fit are 
presented in Eq.~(\ref{eqParameters}).
The comparison with the
spacelike and timelike data~\cite{PionFF_data_TL,PionFF_data_SL}
is presented in Fig.~\ref{fig-FitFpi2}.

One can notice in Fig.~\ref{fig-FitFpi2} 
that the parametrization (\ref{eqFpi}) does not provide 
a high quality description of the $q^2 > 0$ data, 
since there is some underestimation of the data on the left side of the peak.
We recall that, as discussed in Sec.~\ref{secPionCloud},
additional shape parameters are necessary to 
obtain a more accurate description of
the $|F_\pi (q^2)|^2$  data~\cite{PDG-rho,Benayoun93,Pisut68}.

For the present study, it is important to ensure 
that one has an accurate description of the spacelike region 
and that the  parametrization has the 
correct analytic behavior near the two pion threshold, as well
as for very large $Q^2=-q^2$.
Minor deviations on the timelike region 
are not relevant since the meson cloud contributions 
provide in general small corrections to the 
dominant valence quark contributions.

One can now focus on Eq.~(\ref{eqGMpi}). 
Following previous studies on the 
$\gamma^\ast N \to \Delta(1232)$ 
transition~\cite{Timelike2,DecupletDalitz}
we consider a parametrization of the pion cloud 
contribution given by the combination
of the two terms: a term associated with the direct coupling 
with the pion and a term associated with 
the coupling with the intermediate baryon states 
(see Fig.~\ref{figMesonCloud}).

The second term is discussed in the main 
text and includes the function $\tilde G_D$ regulated
by the cutoff $\Lambda_D^2= 0.9$ GeV$^2$
determined in a previous work~\cite{Timelike2}.

The first term includes a tripole factor with a cutoff $\Lambda_\pi^2$.
The cutoff  $\Lambda_\pi^2$ measures the 
falloff of the pion cloud contribution of the diagram (a)
to the magnetic transition form factor.
It can be determined in the case of the 
$\gamma^\ast N \to \Delta(1232)$ transition,   
directly from the comparison with the physical data.
Since the valence quark contribution is well calibrated by the comparison with
the lattice data with heavy quarks, the pion cloud contributions can be determined
by the combination of the two terms of (\ref{eqGMpi})
with the bare contribution, where $\Lambda_\pi^2$ is adjusted to the data. 
We obtain a very good description of the data 
with $\Lambda_\pi^2 = 2.30$ GeV$^2$.
The comparison with the data is presented 
in Fig.~\ref{fig-GM-total}.
Notice that in the figure we consider only
the pion cloud contribution.
The small kaon cloud contribution is not included.

%
%

\setcounter{figure}{0}
\renewcommand\thefigure{\thesection.\arabic{figure}}

\setcounter{table}{0}
\renewcommand{\thetable}{D\arabic{table}}

\section{Meson cloud contributions}
\label{appMesonCloud}

In this appendix, we present the general expressions 
for the meson cloud contributions 
associated with the diagrams (a) and (b) in Fig.~\ref{figMesonCloud}.
The analytic results for the expansion in terms of $Q^2$ 
are presented in Appendix~\ref{appKaonCloud}.

The expressions presented next are derived 
from the cloudy bag model (CBM)
formalism~\cite{Thomas84,Theberge83,Kubodera85,Tsushima88,Yamaguchi89},
where the formulas
are derived in the heavy baryon limit, 
by neglecting the three-momentum of the baryons in the meson cloud loops.

The calculation of the contributions associated 
with the diagrams (a) and (b) from Fig.~\ref{figMesonCloud}
requires the evaluation of integrals of the 
form~\cite{Theberge83,Kubodera85,Tsushima88,Yamaguchi89},
\ba
& &
\hspace{-.35cm}
H_{B'B; B_1}^{M}  =
 \frac{f_{M B'B_1} f_{M B_1B}}{m_\pi^2}  
\int \frac{d^3 {\bf k}}{(2\pi)^3} \; F_{\rm a}(B',B_1,B;M), 
\nonumber \\
& &
\hspace{-.35cm}
H_{B'B; B_2 B_1}^{2M} =  \nonumber \\
& &
\hspace{.6cm}
 \frac{f_{M B'B_2} f_{M B_1B}}{m_\pi^2} 
\int \frac{d^3 {\bf k}}{(2\pi)^3} \; F_{\rm b}(B',B_2,B_1,B;M), 
\nonumber
\ea
where a and b label the CBM integrand functions associated 
with the two diagrams.
The particle labels ($B$, $B'$, $B_1$, $B_2$ and $M$)
include the reference to the 
three-momentum of the particles.

The factors $f_{M B_1 B_2}$ represent the 
 $M B_1 B_2$ couplings determined by the $SU(3)$ symmetry.
The factor $1/m_\pi^2$  is included by convenience
taking into account the ratios between kaon, eta
and pion masses, factored in the functions 
$F_{\rm a}$ and $F_{\rm b}$. 
Using the relations  $f_{M B_1 B_2}$ 
with the  $\pi N N$ coupling  $f \equiv  f_{\pi NN}$,
we can express the product 
of the meson-baryon coupling constants 
in terms of a flavor coefficient 
and the factor $f^2/m_\pi^2$~\cite{Theberge83,Kubodera85,Tsushima88,Yamaguchi89}.
In summary, the integrals can be written as 
a product of a flavor coefficient,  $f^2/m_\pi^2$
and an explicit integral, 
independent of the coupling factors.

To represent the loop integrals, we use the notation, 
\ba
& &
\omega_{\bf  k} = \sqrt{m^2 + {\bf k}^2} \\
& &
\omega_{B_2 B_1} = M_{B_2} - M_{B_1},
\ea
where $m$ is the mass of the meson and 
$M_{B_i}$ is the mass of the baryon $B_i$   ($i=1,2$).
For the case  ${\bf q} \ne {\bf 0}$,
we need also the auxiliary functions,
\ba
\omega_{\pm} = \sqrt{m^2 + \left({\bf k} \pm \sfrac{1}{2}{\bf q} \right)^2},  
\ea
and 
\ba
\rmk_\pm =  |{\bf k} \pm \sfrac{1}{2}{\bf q}|.
\ea

Using the decomposition  in polar coordinates 
with the symmetry in terms of the direction of ${\bf q}$, 
one obtains the relation  
\ba
\int \frac{d^3 {\bf k}}{(2\pi)^3} 
\equiv \frac{1}{4 \pi^2} \int_0^{+ \infty} {\rmk}^2 d {\rmk} 
\int_{-1}^1 dz ,
\ea
where $z= \cos \theta$ and $\rmk \equiv |{\bf k}|$
($\theta$ is the angle between ${\bf k}$ and  ${\bf q}$).
When the integration in $z$ is analytic, 
we can express the final expression in 
terms of the radial variable $\rmk$.

The CBM vertex form factor ${\cal U}({\rmk}  R)$
is represented by~\cite{Thomas84,Theberge83,Kubodera85,Tsushima88,Yamaguchi89}
\ba
{\cal U}( {\rmk} R) = j_0( {\rmk} R) + j_2( {\rmk} R),
\ea
where $j_0$ and $j_2$ are spherical Bessel functions of the first kind.
$R$ is the bag radius ($R \simeq 1$ fm).
In the following, we use $u ={\rmk} R$.

In the present case we need to expand the meson cloud contributions 
up to the first order of $Q^2$ (second order in $u$),
which is sufficient for the present purpose of estimating in the small-$Q^2$ region.
Then, one can use the following expansion for $u ={\rmk} R$:
\ba
{\cal U} \left(\sqrt{u^2 + h} \right) &=& 
{\cal U} (u) + \frac{{\cal U}' (u)}{2u} h \label{eq-expandU1} \\
& &+ 
\frac{1}{4} 
\left[ \frac{{\cal U}^{\prime \prime}(u)}{u^2} - 
 \frac{{\cal U}' (u)}{u^3} 
\right] \frac{h^2}{2} + {\cal O}(h^3). \nonumber 
\ea 
To simplify the final expressions, we use the compact notation,
\ba
{\cal U} \left(\sqrt{ ({\rmk} R)^2 + h}  \right) \simeq 
{\cal U}({\rmk} R)\left[ 1 + g_1 h + g_2 \frac{h^2}{2} 
\right] ,
\ea
where
\ba
& &
g_1 = \frac{1}{2 {\rmk} R} \frac{{\cal U}' ({\rmk} R) }{{\cal U} ({\rmk} R)}, 
\label{eqG1}
\\
& &
g_2 = \frac{1}{4 {\cal U}({\rmk} R)} 
\left[ \frac{{\cal U}^{\prime \prime}( {\rmk} R)}{( {\rmk} R)^2} - 
 \frac{{\cal U}' ( {\rmk} R)}{( {\rmk} R)^3}
\right].
\label{eqG2}
\ea

\begin{table*}[t]
\begin{center}
\begin{tabular}{l  c}
\hline
\hline
    & $G_M^{K{\rm a}}$ \\
\hline
\hline
$\gamma^\ast N \to \Delta$ & 
$\frac{4\sqrt{2}}{9} (2 M_N)  \left[ \frac{1}{25} H^K_\Sigma +  \frac{1}{5} H^K_{\Sigma^\ast} \right]$
\\[.2cm]
$\gamma^\ast \Lambda \to \Sigma^{0*}$ &
$\frac{2\sqrt{2}}{15 \sqrt{3}} (2 M_\Lambda)
\left[ \frac{3}{5} H^K_N -  \frac{1}{5} H^K_{\Xi}  + 2 H^K_{\Xi^\ast}\right]$ \\[.2cm]
$\gamma^\ast \Sigma \to \Sigma^{*}$ &
$\frac{\sqrt{2}}{3} (2 M_\Sigma) \left[
\frac{2}{75} H^K_N +  \frac{8}{15} H^K_\Delta + \frac{2}{15} H^K_\Xi + \frac{4}{15} H^K_{\Xi^\ast}
\right] + $ \\
&   
  $\frac{\sqrt{2}}{3} (2 M_\Sigma) \left[
- \frac{2}{75} H^K_N +  \frac{4}{15} H^K_\Delta + \frac{2}{15} H^K_\Xi + \frac{4}{15} H^K_{\Xi^\ast}
\right] J_3  $ \\[.3cm]
$\gamma^\ast \Xi \to \Xi^{*}$ &  
$\frac{\sqrt{2}}{3} (2 M_\Xi) \left[
-\frac{1}{25} H^K_\Lambda +  \frac{1}{5} H^K_\Sigma + \frac{2}{5} H^K_{\Sigma^\ast} +
\frac{2}{5} H^K_{\Omega}
\right] + $  \\
 & 
$\frac{\sqrt{2}}{3} (2 M_\Xi) \left[
\frac{1}{25} H^K_\Lambda +  \frac{1}{15} H^K_\Sigma + \frac{2}{15} H^K_{\Sigma^\ast} +
\frac{2}{5} H^K_{\Omega}
\right]\tau_3 $ \\[.1cm]
\hline
\hline
\end{tabular}
\end{center}
\caption{\footnotesize Kaon cloud contributions $G_M^{K}$ for the diagram (a).
For simplicity, we use $H^K_{B_1}$ to represent 
$H^K_{B'B;B_1}$, since the states $B'$ and $B$ are evident.
In addition to the results presented above, the proper isospin factors are included 
for the $\Sigma^{* \pm,0}, \Sigma^{\pm,0}, \Xi^{* 0,-}$ 
and $\Xi^{0,-}$ in the final isospin (charge) dependent results 
in Tables~\ref{tableKaonCloud} and~\ref{tableGM0}.} 
\label{table-CoefficientsA}
\end{table*}

\subsection{Diagram (a)}

The contributions from the diagram (a)
for the \mbox{$\gamma^\ast B \to B'$}
can be written as
\ba
G_M^{{\rm MC}a} (Q^2) = \sum_{M,B_1}  C^{M}_{B'B;B_1} H_{B'B; B_1}^{M} (Q^2),
\label{HM-integral}
\ea
where $C^{M}_{B'B; B_1}$ are the coefficients
calculated in the CBM framework.

In Table~\ref{table-CoefficientsA}, we present the 
reduced matrix elements $G_M^{{\rm MC}a}$ for the kaon cloud contributions.
To obtain the contributions of the different components 
of the $\Sigma^* \to \gamma \Sigma$ transitions, 
we replace the operators $J_3$ (isospin 1)
by the charge of the $\Sigma$ ($+,0,-$).
As for the $\Xi^* \to \gamma \Xi$ transitions, 
we replace  $\tau_3$ (isospin 1/2) 
by the $\Xi$ isospin projection 
($\Xi^0$ when $\tau_3 \to +1$; $\Xi^-$ when $\tau_3 \to -1$).

As discussed in the main text, 
the integrals $H_{B'B; B_1}^{M}$ are evaluated in the Breit frame.
We can characterize the diagram (a) with 
the momenta ${\bf k} - \sfrac{1}{2} {\bf q}$
and  ${\bf k} + \sfrac{1}{2} {\bf q}$ for the first 
and second meson lines and a momentum $- {\bf k}$ 
for the  intermediate baryon ($B_1$).
The CBM integrals depend only on the three-momentum 
of the intermediate state particles.
The explicit form is 
\ba
& &H_{B'B; B_1}^{M} (Q^2) = 
\nonumber \\
& &
\frac{f^2}{m_\pi^2}
 \int \frac{d^3 {\bf k}}{(2\pi)^3}
\frac{{\cal U}({\rmk}_+ R) {\cal U}({\rmk}_-R) }{4 \omega_+ \omega_- 
\left[ \left(\omega_+ + \omega_- \right)^2 
- \omega_{B'B}^2 \right]} 
\frac{( \hat {\bf q} \times {\bf k})^2}{4} 
{\cal G}_a 
, \nonumber \\
& &  \label{eqIntA1}
\ea
where 
\ba
{\cal G}_a &= &
\frac{2 \left(\omega_+ + \omega_- \right) 
 \left(\omega_{B_1 B} + \omega_+ + \omega_- \right) +
2 \omega_{B B'} \omega_+ }{
(\omega_{B_1 B'} + \omega_+ )(\omega_{B_1 B} + \omega_-)
} + \nonumber \\
 & &
\frac{2 \left(\omega_+ + \omega_- \right) 
 \left(\omega_{B_1 B} + \omega_+ + \omega_- \right) +
2 \omega_{B B'} \omega_- }{
(\omega_{B_1 B'} + \omega_- )(\omega_{B_1 B} + \omega_+)}. 
\nonumber \\
& &
\ea

In Eq.~(\ref{eqIntA1}), the factor 
$( \hat {\bf q} \times {\bf k})^2 = {\bf k}^2 \sin^2 \theta$
represents the $p$-wave contributions of the mesons 
in the meson cloud loops~\cite{Lu97,Lu98}.
The presence of the last factor 
makes it difficult to take the limit ${\bf q}^2 = 0$.
One needs then to perform the angular integration 
before taking the limit.
The result of the expansion in terms of $Q^2$
is presented in Appendix~\ref{appKaonCloud}.


\begin{table*}[t]
\begin{center}
\begin{tabular}{l  c}
\hline
\hline
    & $G_M^{K{\rm b}}$ \\
\hline
\hline
%
$\gamma^\ast N \to \Delta$ & 
$\frac{2\sqrt{2}}{3} \mu_V  \left[ \frac{4}{25} H^{2K}_{\Sigma \Lambda} 
+  \frac{1}{5} H^{2K}_{\Sigma^\ast \Lambda} +   
\frac{8}{225} H^{2K}_{\Sigma \Sigma} \right. $ \\
& 
$\left.  +  \frac{4}{45} H^{2K}_{\Sigma^\ast \Sigma^\ast} -  
\frac{1}{45} H^{2K}_{\Sigma^\ast \Sigma} 
+ \frac{4}{225} H^{2K}_{\Sigma \Sigma^\ast} 
 \right]$ \\[.35cm]
%
%
$\gamma^\ast \Lambda \to \Sigma^{0*}$ &
$\sqrt{\frac{2}{3}} \frac{M_\Lambda}{M_N} \mu_V
\left[ \frac{4}{15} H^{2K}_{NN} +  \frac{8}{15} H^{2K}_{\Delta N} + \frac{4}{225} H^{2K}_{\Xi \Xi}  \right. $\\
&  $\left.   +
\frac{4}{45} H^{2K}_{\Xi^\ast \Xi} +  \frac{8}{225} H^{2K}_{\Xi \Xi^\ast} + \frac{8}{45} H^{2K}_{\Xi^\ast \Xi^\ast} 
\right]$ \\[.35cm]
%
%
$\gamma^\ast \Sigma \to \Sigma^{*}$ &
$\frac{\sqrt{2}}{3} \frac{M_\Sigma}{M_N} \mu_S
 \left[
\frac{4}{9} H^{2K}_{\Xi^\ast \Xi} -  \frac{8}{225} H^{2K}_{\Xi \Xi^\ast}
\right] + \frac{\sqrt{2}}{3} \frac{M_\Sigma}{M_N} \mu_3 \frac{4}{15} H^{2K}_{\Xi \Xi} + $ \\
&   
 $\frac{\sqrt{2}}{3} \frac{M_\Sigma}{M_N} \mu_V
 \left[
\frac{4}{225} H^{2K}_{N N} +  \frac{16}{45} H^{2K}_{\Delta \Delta} 
\right] 
+ \frac{\sqrt{2}}{3} \frac{M_\Sigma}{M_N} \mu_4 
\frac{8}{45} H^{2K}_{\Xi^\ast \Xi^\ast}   + $ \\[.15cm]
&  $ \frac{\sqrt{2}}{3} \frac{M_\Sigma}{M_N} \mu_V
 \left[ \frac{4}{45} H^{2K}_{NN} - \frac{4}{45} H^{2K}_{\Delta N} +  \frac{16}{225} H^{2K}_{N\Delta} +  \frac{8}{9} H^{2K}_{\Delta \Delta} 
\right. $ \\
& $\left.  + \frac{4}{45} H^{2K}_{\Xi \Xi} + \frac{4}{9} H^{2K}_{\Xi^\ast \Xi} -  \frac{8}{225} H^{2K}_{\Xi \Xi^\ast}
-  \frac{8}{45} H^{2K}_{\Xi^\ast \Xi^\ast}
\right] J_3$ \\[.35cm]
%
%
$\gamma^\ast \Xi \to \Xi^{*}$ &  
$\frac{\sqrt{2}}{3} \frac{M_\Xi}{M_N} \mu_S
  \left[ \frac{2}{3} H^{2K}_{\Sigma^\ast \Sigma}
- \frac{4}{75} H^{2K}_{\Sigma \Sigma^\ast}
\right] 
+  \frac{\sqrt{2}}{3} \frac{M_\Xi}{M_N}
\mu_1 \frac{4}{15} H^{2K}_{\Sigma \Sigma}  + $  \\
  &   
$\frac{\sqrt{2}}{3} \frac{M_\Xi}{M_N} \mu_s
  \left[ \frac{4}{75} H^{2K}_{\Lambda \Lambda}
+ \frac{8}{15} H^{2K}_{\Omega \Omega}
\right] 
+  \frac{\sqrt{2}}{3} \frac{M_\Xi}{M_N}
\mu_2 \frac{8}{45} H^{2K}_{\Sigma^\ast \Sigma^\ast}  + $
\\[.15cm]
 & 
$  \frac{\sqrt{2}}{3} \frac{M_\Xi}{M_N} \mu_V  \left[
 - \frac{4}{75} H^{2K}_{\Sigma \Lambda} +  \frac{2}{15} H^{2K}_{\Sigma^\ast \Lambda}
+ \frac{4}{15} H^{2K}_{\Lambda \Sigma} + \frac{16}{45} H^{2K}_{\Sigma \Sigma} 
\right.$ \\
& 
$\left.
+ \frac{4}{9} H^{2K}_{\Sigma^\ast \Sigma} + \frac{4}{75} H^{2K}_{\Lambda \Sigma^\ast }
- \frac{8}{225} H^{2K}_{\Sigma \Sigma^\ast }
+ \frac{16}{45} H^{2K}_{\Sigma^\ast \Sigma^\ast }
\right]\tau_3 $ \\[.1cm]
\hline
\hline
\end{tabular}
\end{center}
\caption{\footnotesize
Kaon cloud contributions $G_M^{K}$ for the diagram (b).
For simplicity, we use $H^{2K}_{B_2 B_1}$ to represent 
$H^{2K}_{B'B;B_2 B_1}$, since the states $B'$ and $B$ are evident.
In addition to the results presented above, the proper isospin factors are included 
for the $\Sigma^{* \pm,0}, \Sigma^{\pm,0}, \Xi^{* 0,-}$ 
and $\Xi^{0,-}$ in the final isospin (charge) dependent results 
in Tables~\ref{tableKaonCloud} and~\ref{tableGM0}.
The auxiliary factors $\mu_V$, $\mu_S$, 
$\mu_i$ ($i=1,2,3,4$) are defined by Eq.~(\ref{eqMUS}).} 
\label{table-CoefficientsB}
\end{table*}

\subsection{Diagram (b)}

The $\gamma^\ast B \to B'$ meson cloud contributions from the diagram (b)
can be expressed in the form 
\ba
G_M^{{\rm MC}b} (Q^2) = \sum_{M,B_1,B_2}  D^{M}_{B'B;B_2 B_1} H_{B'B; B_2 B_1}^{2M} (Q^2).
\nonumber  \\
& &
\label{HM2-integral}
\ea
The coefficients $D^{M}_{B'B;B_2 B_1}$ are calculated using the CBM.
The results for the reduced matrix elements $G_M^{{\rm MC}b}$
are presented in Table~\ref{table-CoefficientsB}.
We use the same convention to $J_3$ and $\tau_3$
as in Table~\ref{table-CoefficientsA}.

The coefficients $D^{M}_{B'B;B_2 B_1}$ include the 
interaction of the meson with the bare cores,
which depend on valence quark contributions.
Those contributions can be expressed in terms 
of the quark effective magnetic moments 
$\mu_q$, as defined in Refs.~\cite{DecupletDecays,Octet2Decuplet}.
In the exact $SU(2)$ symmetric limit for the meson cloud
($\mu_u= \mu_d$), the coefficients depend only on $\mu_s$
and the average of $u$ and $d$ magnetic moments
\ba
\bar \mu_u = (2\mu_u + \mu_d)/3.
\label{eqMUA}
\ea
To write the final expressions, 
we use the notation, 
\ba 
& &
\mu_S = \frac{1}{3}(\bar \mu_u + 2 \mu_s), \hspace{1cm}
\mu_V = \bar \mu_u, \nonumber  \\
& &
\mu_1 = \frac{1}{3}(2 \bar \mu_u +  \mu_s), \hspace{1cm}
\mu_2 = \bar \mu_u -\mu_s, 
\label{eqMUS}
\\
& &
\mu_3 = \frac{1}{9}(\bar \mu_u +  8 \mu_s), \hspace{1cm}
\mu_4 = \frac{1}{3}(-\bar \mu_u +  4 \mu_s).
\nonumber 
\ea

In this case, we can use a configuration where 
the momentum of $B_1$ is ${\bf k}$,  
the momentum of $B_2$ is ${\bf k + q}$ 
and the momentum of the meson is 
$- \left( {\bf k} + \sfrac{1}{2} {\bf q} \right)$.
The explicit form is 
\ba
& &H_{B'B; B_2 B_1}^{2M} (Q^2) = 
\nonumber \\
& & \frac{f^2}{m_\pi^2}
 \int \frac{d^3 {\bf k}}{(2\pi)^3}
\frac{ {\rmk}_+^2  [{\cal U}({\rmk}_+ R)]^2 }{2 
\omega_+ (\omega_{B_2 B'} + \omega_+ )(\omega_{B_1 B} + \omega_+ )}. 
\nonumber \\
\label{eqIntB1}
\ea

The result of the expansion in terms of $Q^2$
is presented in Appendix~\ref{appKaonCloud}.
For more details about Eqs.~(\ref{HM-integral}) and~(\ref{HM2-integral})
see Ref.~\cite{DecupletDecays}.

\section{Calculation of $G_M^{K\ell}(0)$ and $\frac{d G_M^{K\ell}}{d Q^2} (0)$}
\label{appKaonCloud}

In this appendix, we present the final expressions for 
the kaon cloud contributions in the limit $Q^2=0$, 
$G_M^{K\ell}(0)$, and the result for $\frac{d G_M^{K\ell}}{d Q^2} (0)$.
These results are used to calculate 
the cutoff parameters of the kaon cloud contributions 
associated with  the diagrams (a) and (b),
as described in Sec.~\ref{secKaonCloudP}.
We use here the notation defined in Appendix~\ref{appMesonCloud}.

The kaon cloud contributions, $G_M^{Ka} (Q^2)$
and  $G_M^{Kb} (Q^2)$ for the diagrams (a) and (b) 
are determined by Eqs.~(\ref{HM-integral}) and (\ref{HM2-integral}) for $M=K$.
Since the coefficients $C^K$ and $D^K$ are known, and independent of $Q^2$,
here we need to focus only on the integrals $H^K$ and $H^{2K}$.

\subsection{Diagram (a)}

For small $Q^2$, we decompose the function $H_{B'B;B_1}^{M} (Q^2)$
into two terms
\ba
H_{B'B;B_1}^{K} (Q^2) = H_{B'B;B_1}^{K (0)}  +  H_{B'B;B_1}^{K (1)} Q^2,
\ea
where 
\ba
H_{B'B;B_1}^{K (0)} = H_{B'B;B_1}^{K} (0), \hspace{.3cm}
H_{B'B;B_1}^{K (1)} = \frac{d H_{B'B;B_1}^{K}}{d Q^2} (0).
\nonumber \\
\ea

To write the final results it is convenient to define the variables
\ba
& &
Z_1^a= \frac{\omega_{\bf k}}{\omega_{B_1 B'} + \omega_{\bf k}} 
       \frac{\omega_{\bf k}}{\omega_{B_1 B} + \omega_{\bf k}}, \\
& & 
Z_2^a = 
\left(\frac{\omega_{\bf k}}{\omega_{B_1 B'} + \omega_{\bf k}} \right)^2 
+ \left(\frac{\omega_{\bf k}}{\omega_{B_1 B} + \omega_{\bf k}} \right)^2  - Z_1^a ,
\nonumber \\
& &  \\
& &
Z_3^a= \frac{\omega_{\bf k}}{\omega_{B_1 B'} + \omega_{\bf k}} +  
       \frac{\omega_{\bf k}}{\omega_{B_1 B} + \omega_{\bf k}} ,\\
& &
Z_4^a= \frac{\omega_{B_1 B'} + \omega_{B_1 B}}{
4 \omega_{\bf k} +  \omega_{B_1 B'} + \omega_{B_1 B}}, \\
& &
Z_5^a= \frac{ \omega_{B' B}^2}{
4 \omega_{\bf k} +  \omega_{B_1 B'} + \omega_{B_1 B}} \frac{1}{\omega_{\bf k}} 
Z_1^a \\
& &
a_1=  \frac{ 4 \omega_{\bf k}^2 }{
4 \omega_{\bf k}^2 - \omega_{B'B}^2} + \frac{1}{2} Z_3^a 
+ \frac{1}{2} Z_4^a , \\
& &
a_2= 2 + Z_2^a + Z_5^a
\ea
and the functions
\ba
G_a 
& =& \frac{(4 \omega_{\bf k} +  \omega_{B_1 B'} + \omega_{B_1 B})}{
\omega_{\bf k} 
( \omega_{B_1 B'} + \omega_{\bf k})
( \omega_{B_1 B} + \omega_{\bf k})}  \nonumber \\
& &
\times
\frac{
[{\cal U}({\rmk} R)]^2}{
4 \omega_{\bf k}^2 - \omega_{B'B}^2},  \\
 A & =& \frac{1}{2} g_1 ({\rmk} R)^2 + (g_2 - g_1^2) \frac{({\rmk} R)^4}{5} \nonumber \\ 
 & &-\frac{1}{5}  (1 - a_1 - a_2)\frac{\rmk^4}{4 \omega_{\bf k}^4}
- a_1 \frac{\rmk^2}{4 \omega_{\bf k}^2}.
\ea
The values of $g_1$ and $g_2$ are given by Eqs.~(\ref{eqG1}) and (\ref{eqG2}).

Using the previous notation, we can write
\ba
H_{B'B;B_1}^{K (0)} 
 &=& 
\frac{f^2}{m_\pi^2} \int_{\bf k} \rmk^2 \,G_a , \\ 
H_{B'B;B_1}^{K (1)}
&= &  
\frac{f^2}{m_\pi^2} \int_{\bf k}  A \, G_a,
\ea
where
\ba
\int_{\bf k} \equiv \frac{1}{12 \pi^2} \int_0^{+ \infty}\; \rmk^2 d \rmk.
\label{eq-def-intK}
\ea

To calculate  $G_M^{K{\rm a}}(0)$, 
we need to consider linear combinations of $H_{B'B;B_1}^{K (0)}$ for the 
different intermediate states $B_1$,
according to Eq.~(\ref{HM-integral}).
Similarly, the results for $\frac{d G_M^{K{\rm a}}}{d Q^2} (0)$
are linear combinations of $H_{B'B;B_1}^{K (1)}$ for the 
different intermediate states $B_1$.
The numerical results for $\frac{d G_M^{K{\rm a}} }{d Q^2} (0)/G_M^{K{\rm a}}(0)$
are presented in Table~\ref{tableKaonCloud}.

\subsection{Diagram (b)}

Also for the diagram (b) we can decompose 
the function $H_{B'B; B_2 B_1}^{2K} (Q^2)$ for small $Q^2$, 
\ba
H_{B'B; B_2 B_1}^{2K} (Q^2) = H_{B'B; B_2 B_1}^{2K (0)} + H_{B'B; B_2 B_1}^{2K (1)} Q^2, 
\nonumber \\
\ea
where 
\ba
& & H_{B'B; B_2 B_1}^{2K (0)} = H_{B'B; B_2 B_1}^{2K} (0), \nonumber\\
& &
H_{B'B; B_2 B_1}^{2K (1)} =  \frac{d H_{B'B;B_2 B_1 }^{2K}}{d Q^2} (0).
\label{eqH2-decomp}
\ea

To write the explicit expressions, we use 
\ba
& &
Z_1^b= \frac{\omega_{\bf k}}{\omega_{B_2 B'} + \omega_{\bf k}} 
       \frac{\omega_{\bf k}}{\omega_{B_1 B} + \omega_{\bf k}}, \\
& &
Z_2^b = 
 \left(\frac{\omega_{\bf k}}{\omega_{B_2 B'} + \omega_{\bf k}} \right)^2 
+ \left(\frac{\omega_{\bf k}}{\omega_{B_1 B} + \omega_{\bf k}} \right)^2  + Z_1^b ,
\nonumber \\
& & \\
& &
Z_3^b= 
1 + \frac{\omega_{\bf k}}{\omega_{B_2 B'} + \omega_{\bf k}} + 
       \frac{\omega_{\bf k}}{\omega_{B_1 B} + \omega_{\bf k}}.   
\ea
and 
\ba
G_b&=&
\frac{ [{\cal U}({\rmk} R)]^2 }{2 
\omega_{\bf k} (\omega_{B_2 B'} + \omega_{\bf k} )(\omega_{B_1 B} + \omega_{\bf k})},
\\
B &=& \frac{1}{3}( g_2 + g_1^2) ({\rmk} R)^4 + 
   \frac{1}{6}
   \left( 7 - 2 Z_3^b \frac{\rmk^2}{\omega_{\bf k}^2}  \right) g_1  
   ({\rmk} R)^2  
\nonumber\\
& &   +  \left(\frac{Z_3^b}{2} +  \frac{Z_2^b}{3}  \right)
    \frac{\rmk^4}{4 \omega_{\bf k}^4} 
    - \frac{7}{6} Z_3^b \frac{\rmk^2}{4 \omega_{\bf k}^2} + \frac{1}{4}.
\ea

Using the previous notation, we can write
\ba
 H_{B'B; B_2 B_1}^{2K (0)}  
 &=&  \frac{f^2}{m_\pi^2} \int_{\bf k} {\rmk^2} \, G_b, \\
  H_{B'B; B_2 B_1}^{2K (1)}  
 &=&  \frac{f^2}{m_\pi^2}  \int_{\bf k} B  \, G_b,
\ea
where the integral symbol is defined by Eq.~(\ref{eq-def-intK}).

The final results for $\frac{d G_M^{K{\rm b}} }{d Q^2} (0)/G_M^{K{\rm b}}(0)$
are presented in Table~\ref{tableKaonCloud}.

\subsection{Numerical calculation of $G_M^{K{\rm a}}(0)$ and $G_M^{K{\rm b}}(0)$}

The calculations presented in this work 
are based on the numerical results for 
$G_M^{K{\rm a}}(0)$ and $G_M^{K{\rm b}}(0)$,
and the corresponding derivatives in the limit $Q^2=0$,
presented in this appendix.

Compared to the results obtained using the CBM 
parametrizations, we include a correction on the $f_{\pi NN}$ coupling constant 
given by $f^2 \to {\cal R} f^2$, 
as described below.
This calibration is necessary in order to obtain 
a pion cloud contribution to the $\gamma^\ast N \to \Delta$ transition 
compatible with the experimental data for $G_M (0)$~\cite{DecupletDecays}.

In the covariant spectator quark model, 
the valence quark contribution to the 
magnetic form factor at $Q^2=0$ takes the form~\cite{DecupletDecays}:
\ba
G_M^{\rm B} (0) = \frac{2\sqrt{2}}{3}
\frac{1}{3} (2 \mu_u + \mu_d),
\label{eqGM-CSQM}
\ea
where the quark effective magnetic moment 
can be written as 
\ba
\mu_q = \sqrt{\frac{2}{3}}\left[
\frac{2 M_B}{ M_{B'} + M_B} + \frac{M_B}{M_N} \kappa_q
\right] {\cal I}(0).
\ea
In the previous relation $k_q$ is the quark anomalous magnetic moment 
and ${\cal I}(0)$  is the  overlap integral of the 
radial wave functions associated with the $\gamma^\ast N \to \Delta$ transition 
in the limit $Q^2=0$.

Equation (\ref{eqGM-CSQM}) is consistent 
with the CBM result, assuming the exact $SU(2)$ isospin symmetry, 
and reproduces also the $SU(6)$ naive quark model result
$G_M^B (0) = \frac{2\sqrt{2}}{3} \mu_p$,  
where $\mu_p$ is the proton magnetic moment, 
since $\mu_p$ can also be represented by $\bar \mu_u$
according to Eq.~(\ref{eqMUA})~\cite{Koniuk80,Pascalutsa07,DecupletDecays}.

We used the equivalence between the CBM and the covariant spectator 
quark model results for $G_M^B(0)$ based on 
the effective quark magnetic moment $\mu_q$
to obtain a parametrization of the meson cloud contributions 
which reproduce the experimental data for 
the $\gamma^\ast N \to \Delta$ transition.
The estimate of the covariant spectator quark model 
gives a good description of the data for large $Q^2$ 
(negligible meson cloud effects).
However, to reproduce the low-$Q^2$ data one needs to correct 
the strength of the meson cloud contribution, regulated by $f^2/m_\pi^2$,
in order to obtain the correct result for $Q^2=0$.
This calibration is made by determining the
necessary factor to reduce the CBM estimate to the  covariant spectator
quark model estimate for the 
$\gamma^\ast N \to \Delta$ transition ($G_M^{\rm B} (0) \simeq 1.323$).
One obtains then the factor ${\cal R} = 0.81$~\cite{DecupletDecays}.
Notice, however, that the relative contributions from the  
different meson cloud contributions 
are still determined by the CBM/SU(6) framework.

The values of $G_M^{K{\rm a}}(0)$ and $G_M^{K{\rm b}}(0)$ 
presented in Table~\ref{tableMesonCloud} are determined 
using ${\cal R} = 0.81$.

\end{document}